\documentclass[%
reprint,
superscriptaddress,
prl,
 amsmath,amssymb,
 aps,
]{revtex4-2}

\usepackage{graphicx}
\usepackage{dcolumn}
\usepackage{bm}

\usepackage{amssymb}

\usepackage{xcolor}
\usepackage[normalem]{ulem}

\newcommand{\redwave}[1]{%
  \bgroup
  \markoverwith{\textcolor{red}{\lower3.5pt\hbox{\sixly \char58}}}%
  \ULon{#1}%
  \egroup
}

\usepackage{xr-hyper}

\usepackage{hyperref} 
\usepackage{cleveref}

\usepackage{verbatim}

\usepackage{pdfpages}
\usepackage{pgffor}

\makeatletter
\AtBeginDocument{\let\LS@rot\@undefined}
\makeatother

\begin{document}

\preprint{APS/123-QED}

\title{Emergent Active-Blob Length Scale \\ Governs the Collapse Transition of Tangentially Driven Polymers}

    \author{Baptiste Parage}
\author{Sara Jabbari-Farouji}%
 \email{Contact author: s.jabbarifarouji@uva.nl}
\affiliation{%
Institute for Theoretical Physics, University of Amsterdam, Science Park 904, 1098XH Amsterdam, The Netherlands\\
}%

\date{\today}

\begin{abstract}

The coil-to-globule transition is a paradigmatic phenomenon in polymer physics and biophysics, in which polymer chains evolve from swollen to globular conformations upon decreasing solvent quality. At equilibrium, this transition can be rationalized in terms of a thermal blob length that diverges at the so-called $\Theta$ condition, marking the crossover from self-avoiding to globular conformations. How activity modifies this conformational transition and affects the fundamental length scales at play remains poorly understood. Here, we address this question for tangentially driven polymers using Brownian dynamics simulations, systematically varying both intrachain attraction strength and activity. We find that polar activity smears out the crossover from swollen to globular conformations. Between self-avoiding and $\Theta$ conditions, activity dominates by introducing an emergent folding length scale, which acts as an active blob length. In this regime, we demonstrate that the ratio of thermal to active length scales provides a new scaling variable that leads to a universal master curve for active polymer collapse.

\end{abstract}

\maketitle

Active polymers and filaments, ranging from motor-powered cytoskeletal filaments that drive cellular motion~\cite{conboy2024cytoskeletal,kruse2024actomyosin} to topologically entangled worm collectives that withstand mechanical stresses~\cite{Deblais2020,Deblais2023,Tubificine_worms}, constitute fundamental structural and dynamical elements of living matter. Unlike active rigid bodies, their elongated and flexible shapes result in an inherent coupling between self-propulsion and internal degrees of freedom, creating a rich statistical and nonlinear mechanical behavior that lies beyond the scope of classical polymer physics~\cite{Winkler2020,filament_function}. At the single-filament level, activity can induce either swelling or compaction of purely self-repelling polymers, depending on the mode and strength of active forcing~\cite{Bianco2018, Fazelzadeh2023a,  Malgaretti2025, Malgaretti2025b, Ubertini2024b, Li2023, Anand2018, Anand2020, Buglakov2025}. In three dimensions (3D), studies of polymer collectives have highlighted how activity alters the structural and mechanical properties of polymer melts~\cite{Ubertini2024b,Breoni2025}, or reshapes the isotropic-nematic transition of semiflexible polymers~\cite{Hooijschuur2026}. By contrast, aside from scarce studies~\cite{Das2021, Paul2022a, tripathi2026}, how activity competes with attractive interactions to affect single-chain conformations remains much less understood.

For molecular polymers, changes in solvent quality driven by factors such as temperature or pH, modulate the effective monomer-monomer interactions impacting chain conformation~\cite{deGennes}. Upon decreasing the solvent quality, polymers transition from a swollen-coil conformation to a dense, compact globule, commonly referred to as the coil-to-globule or collapse transition. This fundamental phenomenon provides a minimal physical framework for understanding processes ranging from biomolecular collapse~\cite{Yoshikawa1996} and protein folding~\cite{Ziv2009} to the behavior of stimuli-responsive hyrdogels~\cite{Kojima2018}. Here, we investigate how this paradigmatic transition is modified by tangential polar activity. Specifically, we consider active polymers with tunable intrachain interactions and study the interplay between tangential self-propulsion and attractive interactions in reshaping the coil-to-globule transition, a regime  largely unexplored compared with the self-repelling limit~\cite{Fazelzadeh2023a, Bianco2018, Malgaretti2025, Malgaretti2025b, Ubertini2024b}.

In equilibrium, the coil-to-globule transition arises from the competition between conformational entropy, favoring extended chain configurations, and effective monomer-monomer attraction, favoring compact conformations. At the so-called $\Theta$ condition (\textit{e.g.}, the $\Theta$-temperature), effective two-body interactions vanish and sufficiently long chains exhibit ideal, Gaussian statistics. In the limit of infinite chain length ($N\rightarrow\infty$), the $\Theta$ condition corresponds to a tricritical point separating swollen coils from compact globules \cite{deGennes, deGennes1975}. Sufficiently close to $\Theta$ ($\lvert T - \Theta \rvert / \Theta \ll 1$), polymer conformations are governed by a characteristic thermal length scale, the thermal blob length $\xi_T$, whose value is set by the strength of the two-body excluded-volume parameter $v$. Below this length scale, excluded-volume effects are negligible, and a chain segment containing $g_T$ monomers behaves as an ideal random walk, such that $\xi_T\sim g_T^{1/2}$,  whereas at larger scales net attractive or repulsive interactions determine the global chain conformation. In a good solvent, the net monomer-monomer interaction is repulsive and $\xi_T$ remains finite. The polymer adopts swollen conformations, with its gyration radius following a self-avoiding-walk (SAW), $R_g\sim \xi_T (N/g_T)^{\nu_{\text{SAW}}}$, where $\nu_{\text{SAW}}\approx0.588$. As the $\Theta$ point is approached, $\xi_T$ diverges as $\xi_T\sim \lvert T-\Theta\rvert^{-1}$, such that the polymer becomes ideal on all length scales, with $R_g^{\Theta}\sim N^{1/2}$. In a poor solvent, attractive interactions dominate and $\xi_T$ becomes finite again, driving the chain into a globule with $R_g\sim \xi_T(N/g_T)^{1/3}$. Near $\Theta$, the expansion factor, $\alpha_g \equiv R_g/R_g^{\Theta}$, follows the scaling law $\alpha_g\sim (N/g_T)^{\nu_T-1/2}$, yielding a universal master curve for passive polymer collapse. Deep in the globular regime, higher-order interactions beyond $v$ become increasingly important, causing this simple scaling description to break down~\cite{Wang2014a}. More details are provided in Supplemental Material Sec.~\ref{supp-sec:thermal_blob_picture}~\cite{SM}.

Here, we investigate how polar activity competes with self-attraction in determining chain conformation by  performing Brownian dynamics simulations of  three-dimensional tangentially-driven polymers over a wide range of attraction strengths and activity levels. We find that activity smears out the transition from swollen
to globular conformations while leaving the $\Theta$ point essentially unchanged. Between SAW and $\Theta$ conditions, an activity-induced folding scale $g_a$ emerges, which sets the size of an active blob governing polymer conformational properties. By combining this length scale with the thermal blob picture, we derive a new scaling variable that universally describes the collapse of tangentially driven polymers near the $\Theta$ point.

\begin{figure}[t!]
    \centering

    \includegraphics[width=\linewidth]{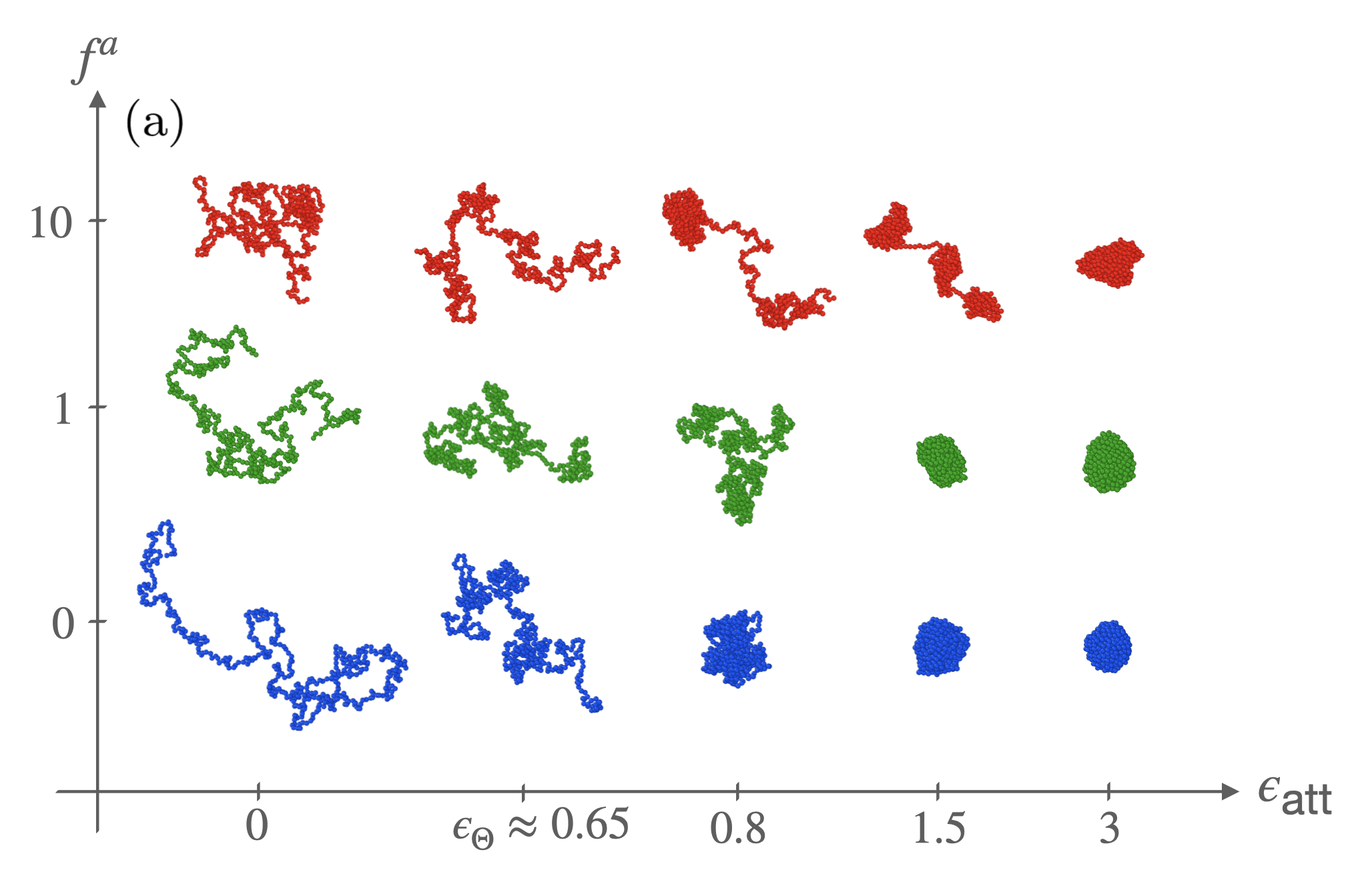}

    \vspace{0.2cm}

    \hspace{-1cm}
    \includegraphics[width=0.8\columnwidth]{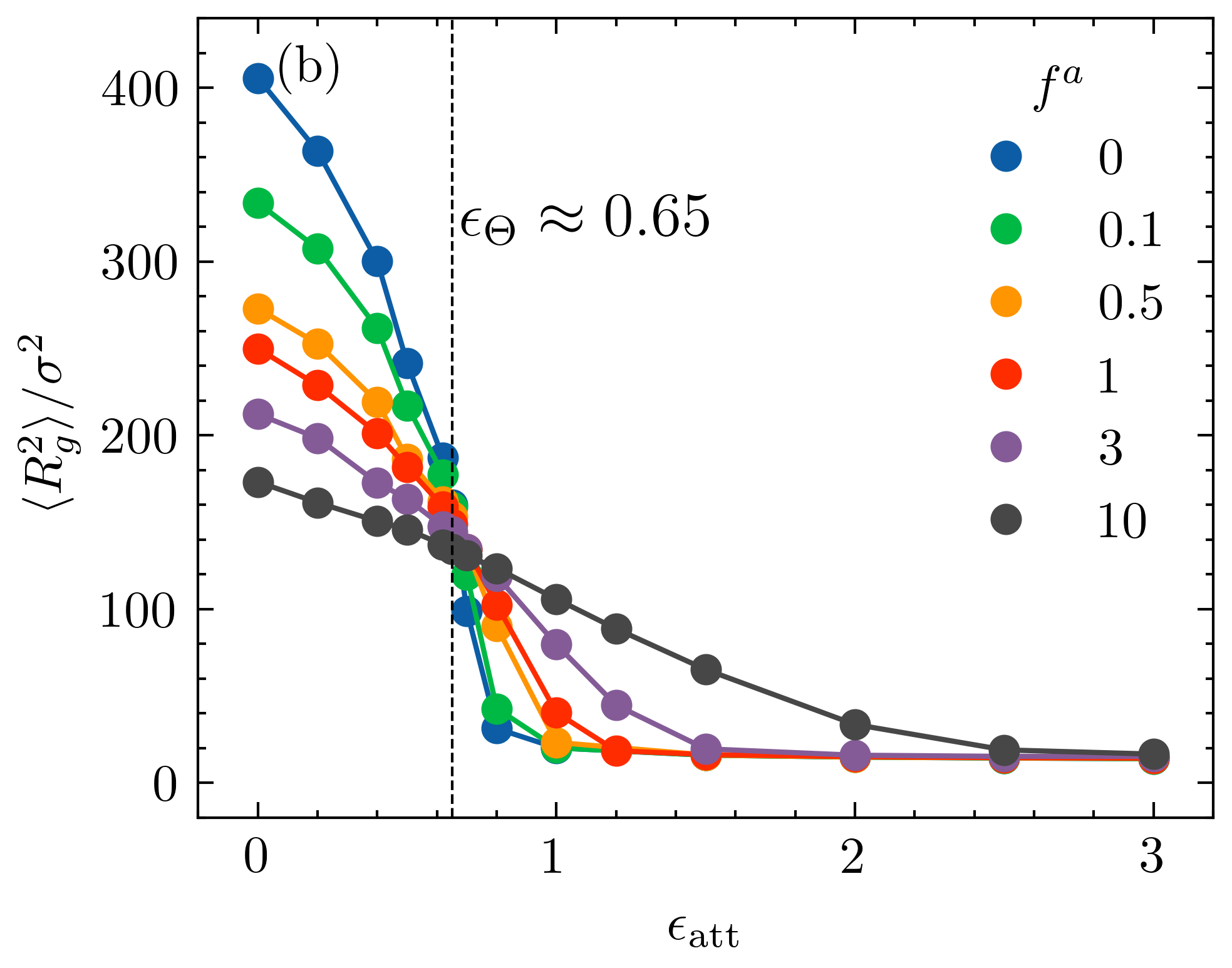}
    
    \caption{\small
    (a) Simulation snapshots of polymer chains of length $N=500$ for active forces $f^{a}=0$, $1$, and $10$ and different attraction strengths $\epsilon_{\text{att}}$. (b) Mean-squared gyration radius, $\langle R_g^2 \rangle$, as a function of the attraction strength $\epsilon_{\text{att}}$ for chain length $N=500$ at different active forces $f^{a}$.}
    \label{fig:snapshots_and_rg}
\end{figure}


 Our simulations consist of a linear, fully flexible bead-spring polymer model of $N$ monomers. Bonded monomers are connected by a harmonic potential with equilibrium bond length $\ell_b,$ chosen to be equal to the bead diameter $\sigma,$ while all pairs of beads interact through excluded-volume interactions described by the repulsive Weeks-Chandler-Andersen (WCA) potential \cite{Weeks1971a}. To model intrachain attractions, we add a short-ranged attractive tail, $U_{\text{att}}(r_{ij})= \frac{1}{2}\epsilon_{\text{att}}\left[\cos(\alpha r_{ij}^2+\beta)-1\right]$ \cite{Soddemann2001a}, which acts between non-adjacent beads. Here, $r_{ij} = \lvert \mathbf{r}_{i,j}\rvert = \lvert \mathbf{r}_{j} - \mathbf{r}_{i}\rvert$ is the interparticle distance and $\epsilon_{\text{att}}$ sets the strength of attractive interactions between polymer beads. 
 A tangential active force is applied to each monomer as $\mathbf{f}^a_i = \frac{f^{a}}{2\ell_b}\left(\mathbf{r}_{i-1,i} + \mathbf{r}_{i,i+1}\right)$. See Supplemental Material Sec.~\ref{supp-sec:polymer_model} \cite{SM} for further details.

We non-dimensionalize our simulations by choosing the monomer diameter $\sigma$ as the unit of length, the thermal energy $k_B T$ as the unit of energy, and $\tau=\gamma \sigma^2/k_B T$ as the unit of time, with $\gamma=1$. The WCA repulsive energy scale $\epsilon_{\text{rep}}$ is fixed to unity, while the attraction strength $\epsilon_{\text{att}}$ is varied in the range $0 \leq \epsilon_{\text{att}} \leq 3$,  spanning both sides of the $\Theta$-point found at $\epsilon_{\text{att}}=\epsilon_{\Theta} \approx 0.65$; see Supplemental Material Sec.~\ref{supp-sec:model_parameters}~\cite{SM}. The active force, 
$f^{a} = f^{a*} k_BT/ \sigma$, is varied in the range $0 \leq f^{a} \leq 20,$ while the chain length is varied between $100 \leq N \leq 500$. The simulations are carried out using HOOMD-blue \cite{Anderson2020a}. See Supplemental Material Sec.~\ref{supp-sec:simulation_parameters} \cite{SM}. 

Snapshots of representative polymer conformations in the steady state are shown in Fig.~\ref{fig:snapshots_and_rg}(a). As $\epsilon_{\text{att}}$ increases, passive polymers ($f^{a}=0$) evolve from swollen coil conformations to compact globules, illustrating the well-known coil-to-globule transition. The same qualitative collapse behavior is observed for active polymers, with two notable features. First, for purely repulsive chains ($\epsilon_{\text{att}}=0$), active polymers appear more compact than their passive counterparts, in agreement with previous findings~\cite{Fazelzadeh2023a, Bianco2018, Malgaretti2025, Malgaretti2025b, Ubertini2024b}. Second, as the active force increases, polymers require larger $\epsilon_{\text{att}}$ to fully collapse. Instead, they develop highly dynamical pearl-necklace-like conformations before eventually collapsing into compact globular states; see Supplemental Material video~\cite{SM}. Further quantitative insight is provided by the mean-squared radius of gyration of the polymer chains, $\langle R_g^2 \rangle$ [Supplemental Material Eq.~\ref{supp-eq:radius_of_gyration}~\cite{SM}], shown in Fig.~\ref{fig:snapshots_and_rg}(b) as a function of $\epsilon_{\text{att}}$ for $N=500$. For passive polymers, $\langle R_g^2 \rangle$ displays a smooth crossover from extended to compact conformations, with an inflection point in the vicinity of $\epsilon_{\Theta}\approx 0.65$. We note that the smooth crossover, rather than a step-like transition, arises from finite $N$, as expected from finite-size scaling predictions~\cite{Miyaki1981, Sun1980, Park1992, Gennes1975, Duenweg2018b}.
Increasing the active force $f^a$ leads to shrinkage of the polymer for $\epsilon_{\text{att}}<\epsilon_{\Theta}$, and to swelling for $\epsilon_{\text{att}} > \epsilon_{\Theta}$, while $\langle R_g^2 \rangle$ remains roughly independent of activity at $\epsilon_{\text{att}}\approx \epsilon_{\Theta}$. A similar behavior is observed for smaller chain lengths, as shown in Supplemental Material Fig.~\ref{supp-fig:rg2_versus_eps_N100_N300}(a-b)~\cite{SM}. Interestingly, increasing activity gradually smears out the transition over a broader range of attraction strengths, an effect very similar to that of decreasing $N$ for passive polymers below $\epsilon_{\Theta}$; see Supplemental Material Fig.~\ref{supp-fig:rg2_versus_eps_N100_N300}(c)~\cite{SM}.

\begin{figure}[t!]
    \centering
    
    \hspace{-0.7cm} 
    \includegraphics[width=1\columnwidth]{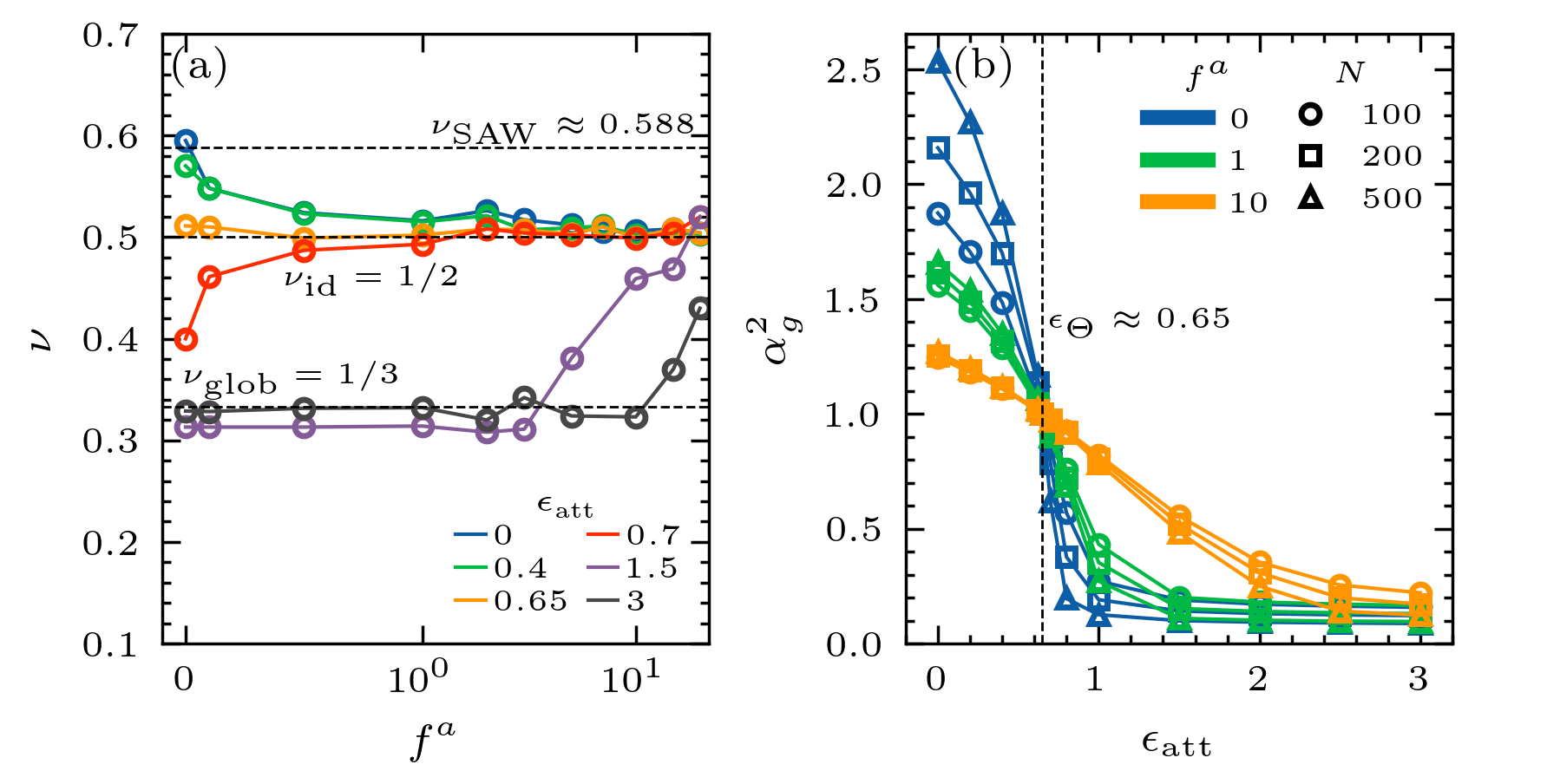}
    
    \caption{\small (a) Global scaling exponent $\nu$ as a function of the active force $f^a$ for different attraction strengths $\epsilon_{\mathrm{att}}$. The $f^a$ axis is linear below unity and logarithmic above. (b) Squared expansion factor $\alpha_g^2 \equiv \langle R_g^2\rangle/\langle R_g^2\rangle_{\Theta}$ versus$\epsilon_{\mathrm{att}}$ for different activities and chain lengths.}
    
    \label{fig:alphag2_and_nu}

\end{figure}

Next, we investigate the scaling behavior of active polymers. For all attraction strengths $\epsilon_{\mathrm{att}}$ and active forces $f^a$, the polymer size follows a power-law dependence on chain length, $\langle R_g^2\rangle \sim N^{2\nu}$, where $\nu(\epsilon_{\mathrm{att}},f^a)$ is the generalized global scaling exponent; see Supplemental Material Fig.~\ref{supp-fig:rg2_scaling_collapse_and_vs_fa}(a)~\cite{SM}. In Fig.~\ref{fig:alphag2_and_nu}(a), we plot $\nu$, obtained from power-law fits of the gyration radius, as a function of the active force $f^a$. In the fully repulsive limit, increasing $f^a$ promotes ideal scaling, with $\nu$ converging toward the value $\nu_{\text{id}}=1/2$, in agreement with earlier findings~\cite{Fazelzadeh2023a, Bianco2018, Malgaretti2025, Malgaretti2025b, Ubertini2024b}. We find that this remains true for $\epsilon_{\text{att}}>0$, where active forcing competes with monomer-monomer attractions. For instance, above $\epsilon_{\Theta}$, the scaling exponent increases from close to the globular value $\nu_{\text{glob}}=1/3$ toward $1/2$, indicating activity-induced swelling into ideal conformations. Thus, large activity effectively promotes $\Theta$-like conditions independently of the strength of intrachain interactions. At the $\Theta$ attraction strength, $\epsilon_{\Theta}\approx0.65$, activity has no effect on the scaling properties of the polymers, which remain ideal, indicating that the $\Theta$ point is unaffected by activity. Similar behavior is observed in the dependence of $\langle R_g^2\rangle$ on the active force $f^a$; see Supplemental Material Fig.~\ref{supp-fig:rg2_scaling_collapse_and_vs_fa}(b-c)~\cite{SM}.

A plot of the squared expansion factor, $\alpha_g^2 \equiv\langle R_g^2\rangle/\langle R_g^2\rangle_\Theta$, as a function of $\epsilon_{\text{att}}$ for different chain lengths $N$ provides insight into the features of the transition [Fig.~\ref{fig:alphag2_and_nu}(b)]. For passive polymers, $\alpha_g^2$ exhibits a sharper transition with increasing $N$,
while the curves for different $N$ intersect at $\epsilon_{\Theta}$. This is expected from the scaling relation $\alpha_g^2\sim (N/g_T)^{2\nu_T-1}$. We find that increasing $f^a$ gradually mitigates the dependence of $\alpha_g^2$ on $N$ for  $\epsilon_{\text{att}} <\epsilon_{\Theta}$, with the curves for different chain lengths becoming progressively closer while still intersecting at $\epsilon_{\Theta}$. This effect suggests that, between SAW and $\Theta$ conditions, the number of thermal blobs in a polymer $N/g_T$, is no longer adequate to describe the collapse. This is supported by Fig.~\ref{supp-fig:alphag_passive_variable}(b-c) in Supplemental Material~\cite{SM}, where the scaling variable $N/g_T$ fails to collapse the expansion factors of active polymers.

\begin{figure}[t!]
    \centering
    \hspace{-0.5cm}
    \includegraphics[width=1\columnwidth]{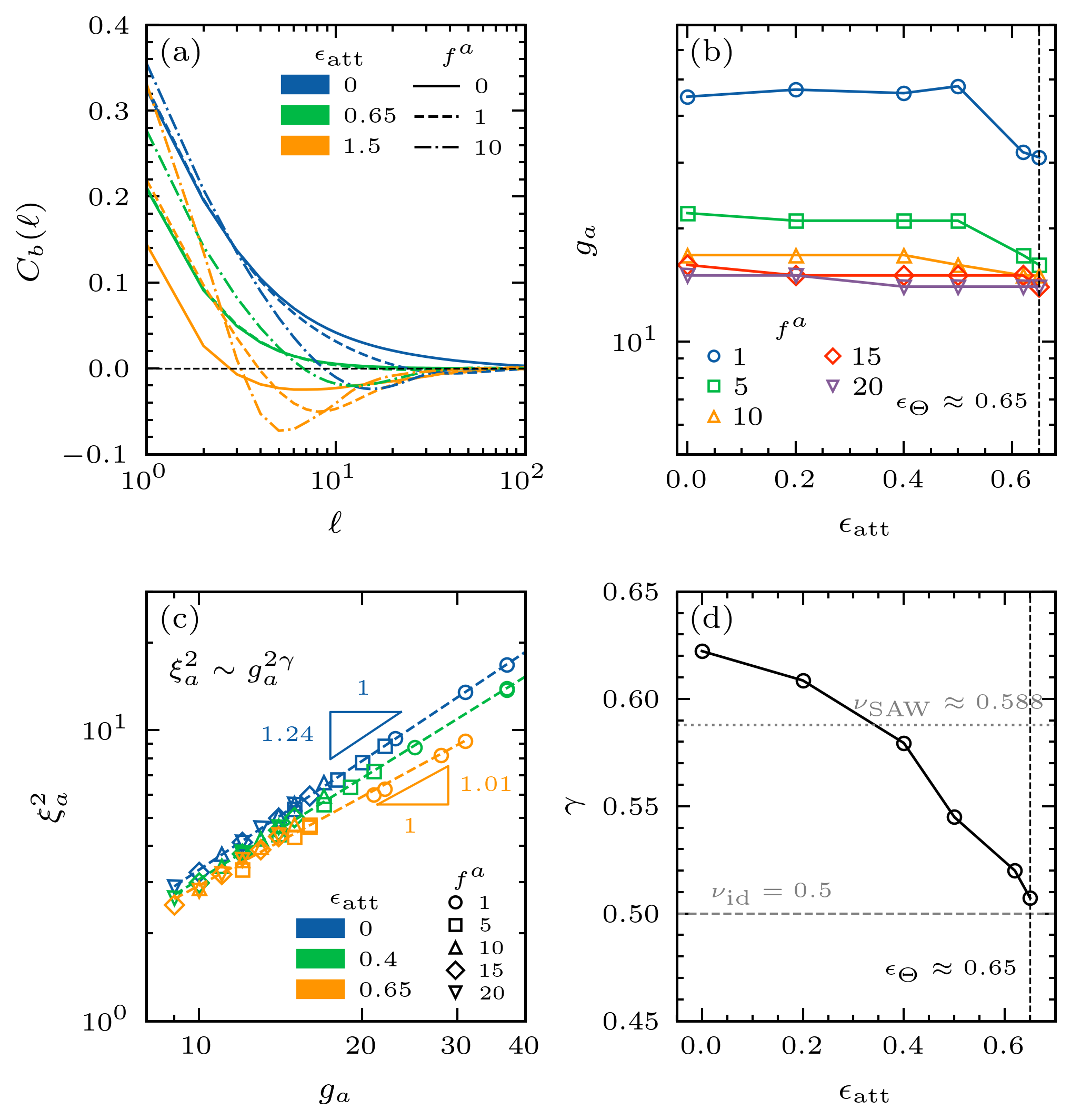}

    \caption{\small
    (a) Bond-bond correlation function $C_b(\ell)$ for different active forces $f^a$ and attraction strengths $\epsilon_{\mathrm{att}}$.
    (b) Active folding length scale $g_a$ versus $\epsilon_{\mathrm{att}}$ for different active forces $f^a$. (c) Active blob size $\xi_a^2=\langle R_g^2(g_a)\rangle$ as a function of $g_a$, using data of all chain lengths. Dashed lines show power-law fits, $\xi_a^2\sim g_a^{2\gamma}$. (d) Exponent $\gamma$ extracted from  the fits in panel (c) against $\epsilon_{\mathrm{att}}$.} 
\label{fig:active_blob_summary}
\end{figure}

To further quantify the effect of activity on chain structure, we compute the bond-bond correlation function, $C_b(\ell)=\langle \mathbf{b}_i\cdot \mathbf{b}_{i+\ell}\rangle$, as shown in Fig.~\ref{fig:active_blob_summary}(a) for different values of attraction $\epsilon_{\text{att}}$ and activity $f^a$. For $\epsilon_{\text{att}} \le \epsilon_{\Theta}$, $C_b(\ell)$ evolves from a monotonically decaying function at $f^a=0$ to one exhibiting a dip at intermediate $\ell$ upon increasing activity. This minimum in bond correlations indicates local polymer backfolding and is a unique signature of tangential activity for $\epsilon_{\text{att}}<\epsilon_{\Theta}$. For $\epsilon_{\text{att}}=1.5$, $C_b(\ell)$ already exhibits a dip in the passive limit, reflecting the attraction-driven collapse into globular structures. We define this folding length scale as $\ell_{\text{min}}=\mathrm{argmin}[C_b(\ell)]$, corresponding to $g_{\min}=\ell_{\text{min}}+2$ monomers. 

We first consider the approach to the $\Theta$ point from the self-repelling  regime. For $\epsilon_{\text{att}}<\epsilon_{\Theta}$, folding is purely induced by activity, and we denote  $g_{\min}\equiv g_a$. We find that $g_a$ is nearly independent of attraction strength and remains larger than the thermal blob size, $g_T \sim 1/v^2$; see Supplemental Material Fig.~\ref{supp-fig:alphag_passive_variable}(a)~\cite{SM}. Beyond curvilinear distances $g_a$, polymer statistics are largely independent of $\epsilon_{\text{att}}$ and display sub-ideal scaling, reflecting large-scale compaction; see Supplemental Material Fig.~\ref{supp-fig:rg2_local_below_theta}(b) showing the scale-dependent gyration radius defined by Supplemental Material Eq.~\ref{supp-eq:local_radius_of_gyration}~\cite{SM}. Below $g_a$, increasing $\epsilon_{\mathrm{att}}$ drives a crossover from self-avoiding to ideal statistics. This is supported by Fig.~\ref{fig:active_blob_summary}(c), where we plot the squared gyration radius at curvilinear distance $g_a$, $\xi_a^2=\langle R_g^2(g_a)\rangle$, as a function of $g_a$ for attraction strengths below $\epsilon_\Theta$. We find a simple power-law dependence, $\xi_a^2\sim g_a^{2\gamma}$, across different activities and chain lengths $N$, where $\gamma$ only depends on attraction strength. Strikingly, the exponent $\gamma$ takes values compatible with the scaling exponent of passive polymers between SAW and $\Theta$ conditions [Fig.~\ref{fig:active_blob_summary}(d)], indicating that segments of length $g_a$ behave as passive, thermal polymers.



\begin{figure}[t!]
    \centering
    \hspace{-1cm}
    \includegraphics[width=1\columnwidth]{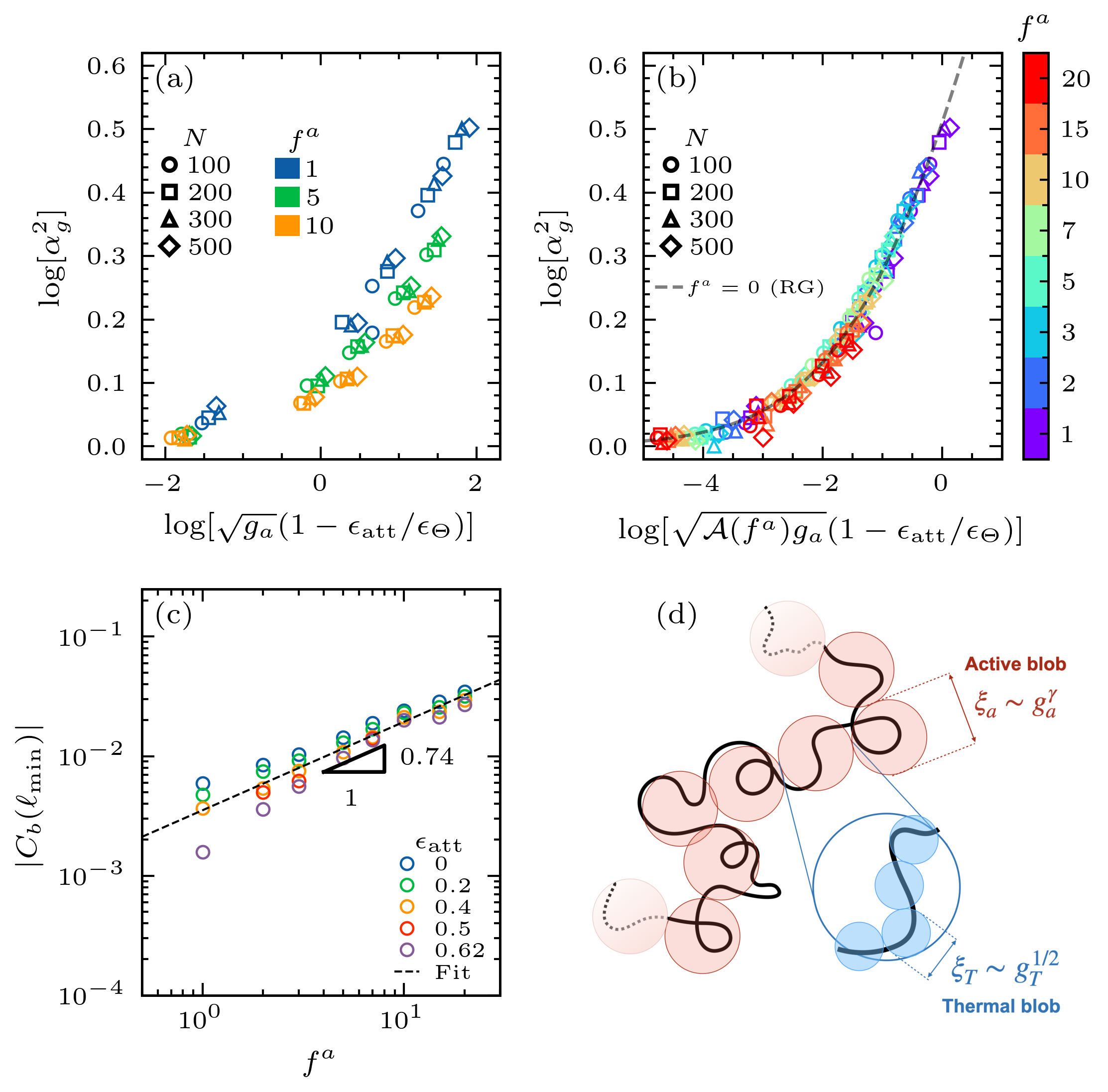}

    \caption{\small
    (a) Squared expansion factor, $\alpha_g^2 \equiv
    \langle R_g^2\rangle/\langle R_g^2\rangle_\Theta$, as a function of the scaling variable $\sqrt{g_a/g_T} \sim\sqrt{g_a}(1-\epsilon_{\mathrm{att}}/\epsilon_\Theta)$
    for different active forces $f^a$ and chain lengths $N$.
    (b) $\alpha_g^2$ against
    $\sqrt{ \mathcal{A}(f^a) \, g_a}(1-\epsilon_{\mathrm{att}}/\epsilon_\Theta)$, with
    $\mathcal{A}(f^a)=k^2({f^a})^{-0.72}$, where $k \approx 0.17$ is the model-dependent constant; see Supplemental Material Sec.~\ref{supp-sec:thermal_blob_picture}~\cite{SM}. The dashed line represents the theoretical expression for $\alpha_g^2$ of passive polymers obtained from renormalization-group (RG) calculations \cite{Schfer2011}, see  Eq.~\ref{supp-eq:fz_rg} in~\cite{SM}].
    (c) Bond anti-correlation strength
$|C_b(\ell_{\mathrm{min}})|$ versus $f^a$
for $\epsilon_{\mathrm{att}}<\epsilon_\Theta$.
(d) Schematic of the active blob picture.}

    \label{fig:active_scaling_and_schematics}

\end{figure}


From the observations above, we make the following thermal blob ansatz for the folding length below $\epsilon_{\Theta}$: $\xi_a^2\sim \xi_T^2(g_a/g_T)^{2\gamma}$. Moreover, we verify that the folding length acts as a blob length scale by plotting $\langle R_g^2\rangle/\xi_a^2$ as a function of $N/g_a$; see Supplemental Material Fig.~\ref{supp-fig:active_blob_ansatz}~\cite{SM}. For $\epsilon_{\text{att}}<\epsilon_{\Theta}$, curves for distinct $f^a$ collapse onto a single master curve, validating the active blob ansatz, $\langle R_g^2\rangle \sim \xi_a^2(N/g_a)^{2\nu_a}$. Thus, the expansion factor of tangentially driven active polymers scales as $\alpha_g^2\sim(g_a/g_T)^{2\gamma-1}$, as derived in Supplemental Material Sec.~\ref{supp-sec:active_blob}~\cite{SM}. Here, we expect finite $N$ effects to modify this scaling to a crossover scaling function, $\alpha_g^2\sim F_a(g_a/g_T)$, where $F_a$ depends on the active force $f^a$. This prediction is verified in Fig.~\ref{fig:active_scaling_and_schematics}(a), where $\alpha_g^2$ for different $N$ collapses onto activity-dependent master curves. The failure of $N/g_T$ as a suitable scaling variable for active polymers is attributed to the weak dependence of $g_a$ on $N$, as shown in Supplemental Material Fig.~\ref{supp-fig:ga_versus_N}(b-c). 
We find that multiplying  the  scaling variable $g_a/g_T  \sim g_a (1-\epsilon_{\mathrm{att}}/\epsilon_\Theta)^2$ by an empirical activity-dependent factor, $\mathcal{A}(f^a)\sim(f^a)^{-0.72}$, collapses all expansion-factor data onto the crossover master curve for passive polymers, as shown in Fig.~\ref{fig:active_scaling_and_schematics}(b). This factor nearly compensates for the activity-induced increase in the strength of bond anti-correlations, $ \lvert C_b(\ell_{\min})\rvert\sim(f^a)^{0.74}$ presented in Fig.~\ref{fig:active_scaling_and_schematics}(c). The increasingly negative bond anti-correlations indicate stronger chain backfolding, resulting in more compact conformations with increasing activity. Consistently, rescaling $g_a/g_T$ by the magnitude of bond correlations itself collapses the data for different $f^a$ well onto the passive master curve as well; see Supplemental Material Fig.~\ref{supp-fig:alpha_g_cb_scaling}~\cite{SM}. Figure~\ref{fig:active_scaling_and_schematics}(d) summarizes our active-blob picture for $\epsilon_{\mathrm{att}}<\epsilon_{\Theta}$, in which active blobs of size $g_a$ consist of smaller thermal blobs of size $g_T$.

\begin{figure}[t!]
    \centering
   \vspace{0.3cm}
   \includegraphics[width=1\columnwidth]{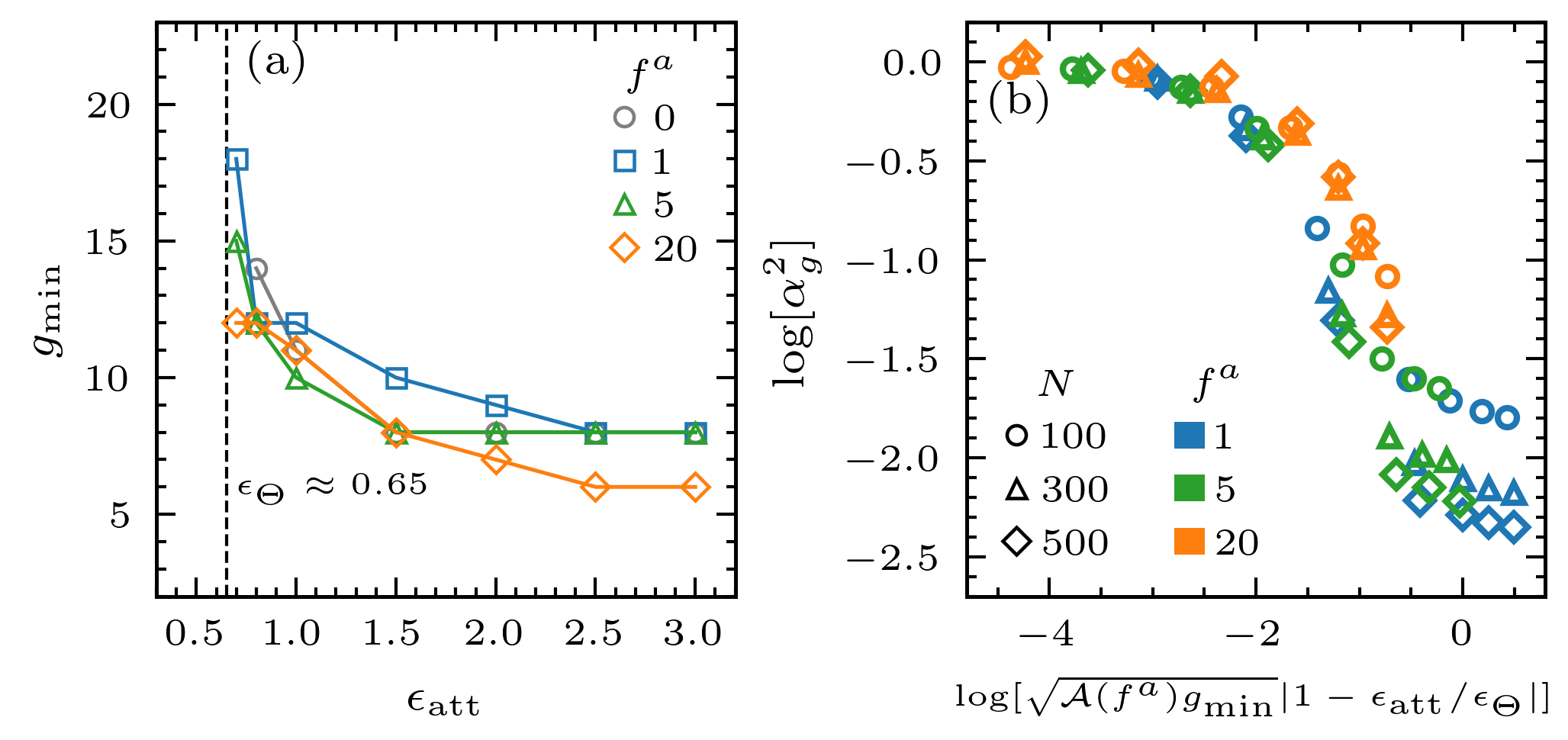}
    
    \caption{\small (a) Folding length scale $g_{\text{min}}=\mathrm{argmin}[C_b(\ell)] +2$, as a function of $\epsilon_{\mathrm{att}}$ beyond $\epsilon_{\Theta}$ for different active forces $f^a$. (b) Squared expansion factor, $\alpha_g^2 \equiv \langle R_g^2\rangle/\langle R_g^2\rangle_\Theta$, as a function of the scaling variable $\sqrt{\mathcal{A}(f^a) \, g_{\text{min}}}|1-\epsilon_{\mathrm{att}}/\epsilon_\Theta|$, with $\mathcal{A}(f^a)=k^2{f^a}^{-0.72}$, for active forces $f^a=1$, 5, and 20 and different chain lengths.} 
    
\label{fig:gmin_and_alphag}
\end{figure}

Next, we focus on the other side of the transition where we move from the $\Theta$ condition to the globular regime.
For $\epsilon_{\mathrm{att}} >\epsilon_{\Theta}$, local folding is no longer a unique signature of activity, as strong attractive forces drive the formation of monomer clusters along the chain [Fig.~\ref{fig:snapshots_and_rg}(a)]. Thus, one cannot unambiguously separate thermal and active contributions to the minimum of the bond-bond correlation function. Figure~\ref{fig:gmin_and_alphag}(a) shows that passive and active polymers exhibit comparable values of $g_{\min}$, while active polymers remain comparatively more swollen than their passive counterparts; see Fig.~\ref{fig:snapshots_and_rg}(a) and Supplemental Material Fig.~\ref{supp-fig:rg2_local_beyond_theta}.  Additionally, we find $g_{\text{min}} > g_T$, with $g_T \sim 1/v^2$; see Supplemental Material Fig.~\ref{supp-fig:alphag_passive_variable}(a)~\cite{SM}. In Fig~\ref{fig:gmin_and_alphag}(b), we show $\alpha_g^2$ as a function of $g_{\text{min}}/g_T$ beyond $\epsilon_{\Theta}$.
 Near the $\Theta$ point, where active chains remain relatively swollen,  the curves for different $N$ collapse well.  At $\epsilon_{\text{att}} \gg \epsilon_{\Theta}$, systematic $N$-dependent deviations emerge, with their onset shifting to larger $\epsilon_{\mathrm{att}}$ with $f^a$. Thus, while activity maintains chains sufficiently swollen and suppresses  higher-order interactions, $g_{\text{min}}/g_T$ remains the relevant scaling variable.


Deep in the globular regime, the local statistics of active polymers closely resemble those of passive chains, see Supplemental Material Fig.~\ref{supp-fig:rg2_local_beyond_theta}, and $g_{\text{min}}$ marks the crossover from ideal statistics at short length scales to globular statistics at larger scales. Further evidence that activity does not alter the internal structure of globules is provided by contact probabilities $p_c(s)$ that are independent of $f^a$ in globular conformations; see Supplemental Material Fig.~\ref{supp-fig:contact_properties}(b)~\cite{SM}. This suggests that a collapsed-blob description based on screened interactions should remain meaningful. Given the strong finite $N$ in the globular regime~\cite{Wang2014b}, we defer a detailed analysis of active globules to future work. 

\begin{figure}[t!]
    \centering
    \hspace{-0.6cm}
    \includegraphics[width=1.05\columnwidth]{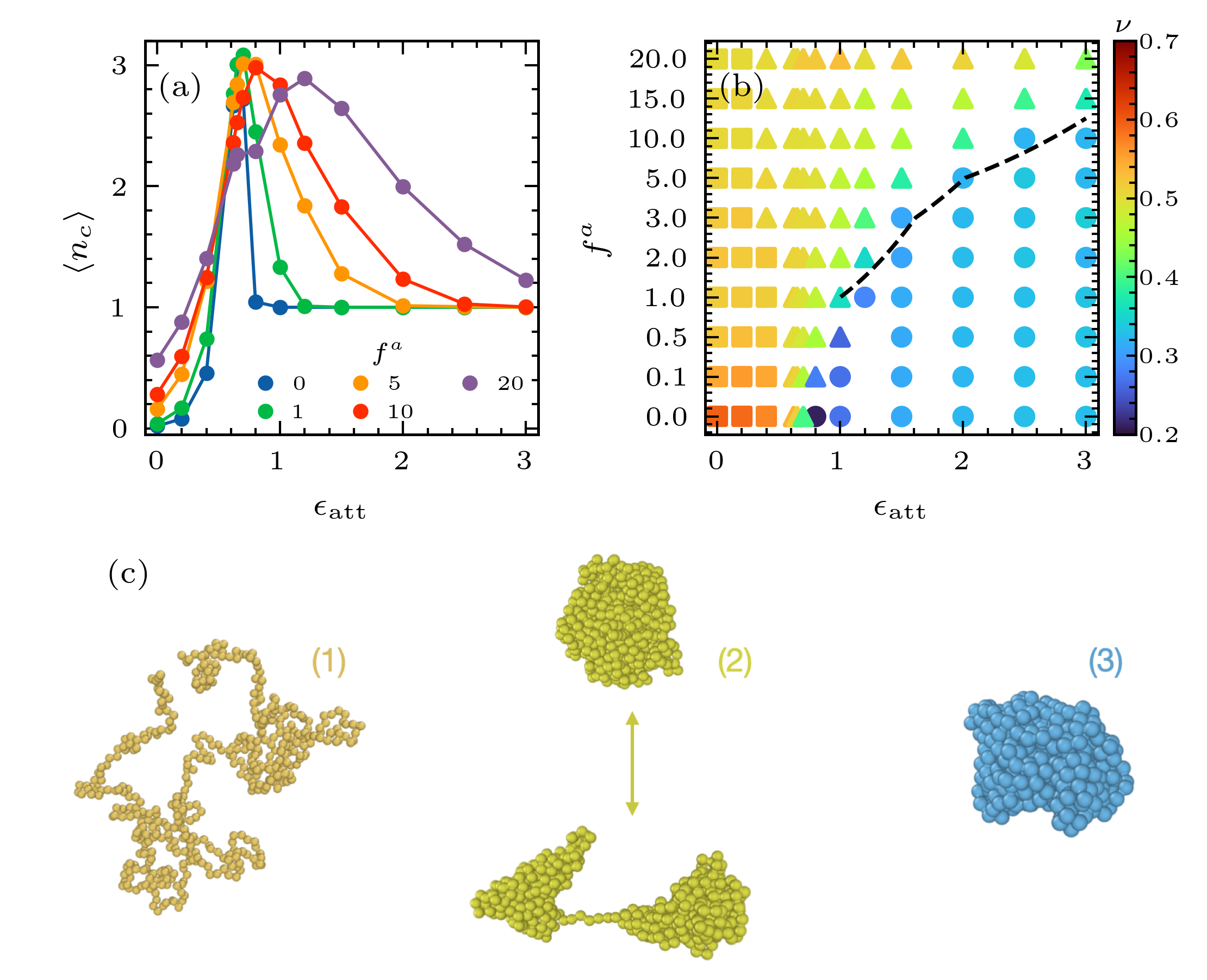}  
    \caption{\small
(a) Average number of clusters along a chain $\langle n_c \rangle$ versus attraction strength $\epsilon_{\mathrm{att}}$
for different active forces $f^a$ at $N=500$.
(b) Steady-state diagram as a function of $\epsilon_{\mathrm{att}}$
and $f^a$ for $N=500$.
Symbols indicate coil ($\square$), necklace ($\triangle$),
and globule ($\circ$) states, classified using $\langle n_c \rangle$.
The black dashed line indicates $f^a \sim \epsilon_{\mathrm{att}}^{2.3}$. See Supplemental Material Sec.~\ref{supp-sec:cluster_analysis}~\cite{SM} for details on the classification.
}

    \label{fig:clustering_and_pd}
    
\end{figure}

Because the validity of the active blob picture depends on chain conformation, we next map the conformational behavior across attraction strengths and active forces. We compute the average number of monomer clusters along a chain, $\langle n_c \rangle$~\cite{Paul2022}; see Supplemental Material Sec.~\ref{supp-sec:cluster_analysis}~\cite{SM}.  Figure~\ref{fig:clustering_and_pd}(a) shows  $\langle n_c \rangle$ as a function of $\epsilon_{\text{att}}$. 
Interestingly, $\langle n_c\rangle$ rises from zero in the repulsive limit to a peak of approximately three near $\epsilon_\Theta$, before decreasing to unity at strong attraction strengths, corresponding to a single globule. Increasing $f^a$ broadens the crossover to the globular state. In this region, active chains exhibit transient pearl-necklace-like structures, intermittently switching between configurations with different cluster numbers; see Supplemental Material video~\cite{SM}. Activity enlarges the  stability range of these clustered states, delaying the collapse into a globular structure.

 In Fig.~\ref{fig:clustering_and_pd}(b), we summarize the steady-state conformations of active polymers as a function of attraction strength $\epsilon_{\mathrm{att}}$ and active force $f^a$, using the mean cluster number $\langle n_c\rangle$ to characterize clustering and color to indicate the global scaling exponent $\nu(\epsilon_{\mathrm{att}},f^a)$. Tangential activity promotes clustered conformations and reduces the region exhibiting SAW scaling, $\nu\approx\nu_{\mathrm{SAW}}$, while broadening the regime of ideal scaling, $\nu\approx\nu_{\mathrm{id}}=1/2$. This extended $\Theta$-like region shows that solvent quality alone no longer determines the large-scale conformational statistics. Conversely, globular conformations persist at strong attraction, but their onset shifts to larger $\epsilon_{\mathrm{att}}$ as activity increases.

In conclusion, by tuning attractive intrachain interactions in tangentially driven active polymers, we demonstrate that  activity smooths out the crossover from swollen to compact conformations.  We rationalize this behavior by extending the thermal blob concept to an active blob picture. Between purely repulsive and $\Theta$ conditions, we identify a characteristic curvilinear distance between monomers $g_a$, nearly independent of attraction strength, that defines the active blob and characterizes chain backfolding into a structure with ideal statistics at larger scales. The size of the active blob,  $\xi_a \sim  g_a ^{\gamma(\epsilon_{\text{att}})}$, is larger than the thermal blob size $\xi_T$ and behave like a passive thermal polymer. 
Combining active and thermal blob scales, we constructed an activity-dependent scaling variable that collapses the expansion factor onto a universal master curve across a wide range of activities near the $\Theta$ point. More generally, within the framework of second-order phase transitions, activity acts as if it modified the effective system size \cite{Cardy1996}.  
The active blob concept provides a basis for extending polymer scaling arguments to various self-driven chain systems and exploring its applicability to semidilute solutions~\cite{Jain2012a,Prakash2019}, confined polymers~\cite{Nikoofard2014}, polymer brushes~\cite{Halperin1994}, stretching~\cite{Pincus1976}, and translocation~\cite{Sakaue2007}.

We acknowledge A. Deblais for fruitful discussions. Computational resources were provided by the Dutch National e-Infrastructure, supported by SURF Cooperative, and the University of Amsterdam Science Faculty (UvA/FNWI) High Performance Computing Facility.

\bibliography{single_chain_PRL}




\section*{Supplementary material}

# Supplementary Materials for: Emergent Active-Blob Length Scale Governs the Collapse Transition of Tangentially Driven Polymers

Baptiste Parage[1] and Sara Jabbari-Farouji[1, *]

[1] *Institute for Theoretical Physics, University of Amsterdam, Science Park 904, 1098XH Amsterdam, The Netherlands.*



## I. SIMULATION METHOD

### A. Polymer model

We model a polymer chain as $N$ connected spherical monomers of diameter $\sigma$ and mass $m$. The dynamics of each monomer $i \in [1, N]$ along the chain is described by the Langevin equation in three-dimensions

$$m\ddot{\mathbf{r}}_i = -\gamma\dot{\mathbf{r}}_i - \sum_{j\neq i} \nabla_i U(r_{ij}) + \mathbf{f}_i^a + \mathbf{f}_i^r, \tag{S1}$$

where $\mathbf{r}_i$ is the position of monomer $i$, $r_{ij} = |\mathbf{r}_{i,j}| = |\mathbf{r}_i - \mathbf{r}_j|$ is the distance between monomers $i$ and $j$, $\gamma$ is the friction coefficient of a bead with the surrounding medium, $U(r_{ij})$ is the potential energy, and $\mathbf{f}_i^a$ and $\mathbf{f}_i^r$ are the active and random forces acting on bead $i$, respectively. The potential energy $U(r_{ij})$ includes contributions from bonded and non-bonded monomers within the same chain. Bonded interactions between adjacent monomers are modeled using a harmonic potential

$$U_{\text{bond}}(r_{ij}) = \frac{k_s}{2}(r_{ij} - \ell_b)^2 \tag{S2}$$

where $k_s$ and $\ell_b$ denote the spring stiffness and bond equilibrium length, respectively, and $|i - j| = 1$. All pair of monomers interact through excluded volume interactions, modeled using a Weeks-Chandler-Anderson (WCA) potential [1] ,

$$U_{\text{WCA}}(r_{ij}) = 4\epsilon_{\text{rep}} \left[ \left(\frac{\sigma}{r_{ij}}\right)^{12} - \left(\frac{\sigma}{r_{ij}}\right)^{6} + \frac{1}{4} \right] \tag{S3}$$

when $r_{ij} \leq 2^{1/6}\sigma$ and 0 otherwise, and where $\epsilon_{\text{rep}}$ is the pairwise repulsive interaction strength. The effect of varying solvent quality is accounted for by adding an short-ranged attractive tail to Eq. S3 for non-adjacent monomers,

$$U_{\text{att}}(r_{ij}) = \frac{1}{2}\epsilon_{\text{att}} \left[\cos\left(\alpha r_{ij}^2 + \beta\right) - 1\right] \tag{S4}$$

for $2^{1/6}\sigma \leq r_{ij} < r_c$. Parameters $\alpha$ and $\beta$ are determined by demanding continuity with Eq. S3 at $r = 2^{1/6}\sigma$, and by imposing that $U_{\text{att}}$ is zero at the cut-off value $r_c$. We choose $r_c = 1.5\sigma$ to include only the first neighboring shell of interactions, ensuring that the potential decays rapidly to zero and thus increases simulation efficiency [2]. Following this choice, one finds $\alpha = 3.1730728678$ and $\beta = -0.856228645$. Together, Eqs. S3 and S4 form the Soddemann-Dunweg-Kremer (SDK) potential [2]

$$U_{\text{SDK}} = \begin{cases} U_{\text{WCA}}(r_{ij}) - \epsilon_{\text{att}} & r_{ij} < 2^{1/6}\sigma \\ U_{\text{att}}(r_{ij}) & 2^{1/6}\sigma \leq r_{ij} < 1.5\sigma \\ 0 & r_{ij} \geq 1.5\sigma \end{cases} \tag{S5}$$

* Correspondence to: s.jabbarifarouji@uva.nl

The active force $\mathbf{f}_i^a$ acting on inner momomers $i \in [2, N-1]$ is given by the tangential model

$$\mathbf{f}_i^a = \frac{f^a}{2\ell_b}\left(\mathbf{r}_{i-1,i} + \mathbf{r}_{i,i+1}\right) \tag{S6}$$

with $f^a$ the active force magnitude. For the end-monomers $i = 1$ and $N$, the active force is defined as

$$\mathbf{f}_1^a = \frac{f^a}{2\ell_b}\mathbf{r}_{1,2} \tag{S7}$$

$$\mathbf{f}_N^a = \frac{f^a}{2\ell_b}\mathbf{r}_{N-1,N} \tag{S8}$$

The random force $\mathbf{f}_i^r$ accounting for thermal fluctuations is modeled by gaussian white noise with zero-mean, $\langle \mathbf{f}_i^r \rangle = \mathbf{0}$, and $\delta$-correlations, $\langle \mathbf{f}_i^r(t) \cdot \mathbf{f}_i^r(t') \rangle = 6k_BT\gamma\delta_{ij}\delta(t'-t)$, where $k_B$ is the Boltzmann constant and $T$ the temperature.

### B. Units and simulation parameters

We choose $\sigma$ as the unit of length, $k_BT$ as the unit of energy, and $\tau = \gamma\sigma^2/\epsilon$ as the unit of time, with $\gamma = 1$. The repulsive strength in Eq. S3, $\epsilon_{\text{rep}}$, is set to 1, while the attractive strength in Eq. S4, $\epsilon_{\text{att}}$, is varied between 0 and $3k_BT$. The equilibrium bond length defined in Eq. S2 is set to $\ell_b = \sigma$, with a spring stiffness of $k_s = 5000k_BT/\sigma^2$ to ensure that the bond length remains unchanged even at large activities [3]. In this way, the contour length of a polymer is simply $L = (N-1)\sigma$. The dimensionless active force, $f^a\sigma/k_BT$, is identical to the commonly used Péclet number [4].

Active polymers are simulated in the overdamped limit, however using the Langevin integrator which allows for a larger timestep. To ensure inertia is negligible in Eq. S1, the inertial timescale of a monomer must satisfy $\tau_m = m/\gamma \ll 1$. Additionally, in the case of active units, $\tau_m$ should be much smaller than the active advection timescale of a monomer, $\tau_a = \sigma\gamma/f^a$ [3]. In our simulations, we set $m = 0.01(\sigma\gamma)^2/k_BT$ and the time step $\delta t = 10^{-3}\tau$ for $f^a \leq 10$, and $m = 0.001(\sigma\gamma)^2/k_BT$ and $\delta t = 10^{-4}\tau$ for $f^a > 10$. Systems comprise between 39 and 234 chains, depending on the parameters.

## II. THERMAL BLOB PICTURE OF THE COLLAPSE TRANSITION

The coil-to-globule or collapse transition of polymers falls within the phenomenology of second-order phase transitions [5], with the solvent quality as the control parameter and the typical polymer size $R$, here quantified by the radius of gyration $R_g$ [Eq. S18], as the order parameter. In the limit of infinite chain length $N \to \infty$, the critical point of the transition corresponds to a tricritical point, conventionally referred to as the $\Theta$ point (*e.g.*, the $\Theta$ temperature). If the interactions between the polymer and solvent molecules are simple and Lennard-Jones-like, the system exhibits an upper critical solution temperature (UCST), and the solvent quality decreases upon decreasing $T$. For $T > \Theta$, polymer chains are swollen, whereas for $T < \Theta$, they collapse into globular conformations. At $T = \Theta$, effective two-body interactions vanish, and the polymer behaves as an ideal chain. This behavior can be summarized by the step-like, temperature-dependent power-law scaling of $R_g$ with $N$,

$$R_g \sim N^{\nu_T} \quad \text{with} \quad \nu_T = \begin{cases} \nu_{\text{SAW}} \approx 0.588 & \text{for } T > \Theta, \\ \nu_{\text{id}} = 1/2 & \text{for } T = \Theta, \\ \nu_{\text{glob}} = 1/3 & \text{for } T < \Theta, \end{cases} \tag{S9}$$

where SAW, id, and glob denote self-avoiding-walk, ideal, and globular conformations, respectively. It is noteworthy that for more complex polymer-solvent interactions, such as hydrogen bonding, decreasing $T$ may increase the solvent quality, resulting in a lower critical solution temperature (LCST). In the following discussion, we focus on the UCST case. The LCST case follows analogously upon reversing the temperature dependence.

Sufficiently close to $\Theta$, typically $|T-\Theta|/\Theta \ll 1$, the conformations of a polymer chain are controlled by a fundamental length scale $\xi_T$, commonly referred to as the thermal blob size. At length scales smaller than $\xi_T$, excluded-volume effects are negligible and chain segments are locally ideal, such that $\xi_T = \ell_K g_T^{1/2}$, where $\ell_K$ is the Kuhn length and $g_T$ is the corresponding number of monomers within a so-called thermal blob. The polymer size can therefore be coarse-grained in terms of thermal blobs,

$$R_g \sim \xi_T \left(\frac{N}{g_T}\right)^{\nu_T} \sim g_T^{1/2-\nu_T} N^{\nu_T}, \tag{S10}$$

where $N/g_T$ is the number of thermal blobs contained in a polymer of $N$ monomers. The quantity $g_T$ is typically defined considering the pairwise monomer interaction energy is of order $k_BT$ [5], leading to

$$\frac{vg_T^2}{\xi_T^3} \sim k_BT \Leftrightarrow g_T \sim k_BT\frac{\ell_K}{v^2} \tag{S11}$$

where $v = \int \mathrm{d}\mathbf{r}\left[1 - e^{-U(\mathbf{r})/k_BT}\right]$ is the excluded-volume parameter, with $U$ the interaction potential between two monomers. At $T = \Theta$, $v = 0$ and close to $\Theta$, $v \approx \partial_T v_\Theta (T - \Theta)$ to first order. Thus, in the vicinity of $\Theta$, the thermal blob variables diverge as

$$g_T \sim |T - \Theta|^{-2}, \qquad \xi_T \sim |T - \Theta|^{-1}, \tag{S12}$$

and the number of thermal blobs scales as

$$\frac{N}{g_T} = k^2 N \left|1 - \frac{T}{\Theta}\right|^2 \tag{S13}$$

where $k$ is the so-called chemistry constant. An estimate of the blob length scale $g_T$ is shown Fig. S1(a). The ratio of the polymer size at $T$ to that at $\Theta$, or expansion factor $\alpha_g \equiv R_g/R_g^\Theta$, scales as

$$\alpha_g \sim \left(\frac{N}{g_T}\right)^{\nu_T - 1/2}, \quad \text{or} \quad \alpha_g \sim z^{2\nu_T - 1}, \tag{S14}$$

where $z = \sqrt{N/g_T} = k\sqrt{N}\,|1 - T/\Theta|$ can thus be identified as the universal scaling variable governing the collapse transition. For finite $N$, Eq. S14 generalizes to a smooth, universal scaling function

$$\alpha_g \sim F(z), \tag{S15}$$

with asymptotic limits $F \sim z^{2\nu_T - 1}$ for $z \gg 1$ and $F \sim 1$ for $z \ll 1$. For $T > \Theta$, renormalization-group calculations suggest an expression of the form [6]

$$F(z) = \left(1 + az + bz^2 + cz^3\right)^m, \tag{S16}$$

which, when fitted to values of $\alpha_g$ obtained from Brownian dynamics simulations of linear polymers, yields $a = 9.528$, $b = 19.48 \pm 1.28$, $c = 14.92 \pm 0.93$, and $m = 0.133913 \pm 0.0006$ [7]. In Fig. S1(b), we successfully verify that the expansion factor from our simulation data of passive polymers ($f^a = 0$) between SAW and $\Theta$ conditions collapses onto Eq. S16 when plotted as a function of $z$. On the globular side of the $\Theta$ point [Fig. S1(c)], we observe a systematic shift of $\alpha_g$ with chain length $N$ deep in the globular phase, as expected from higher-order interactions and strong finite $N$ effects [8, 9].

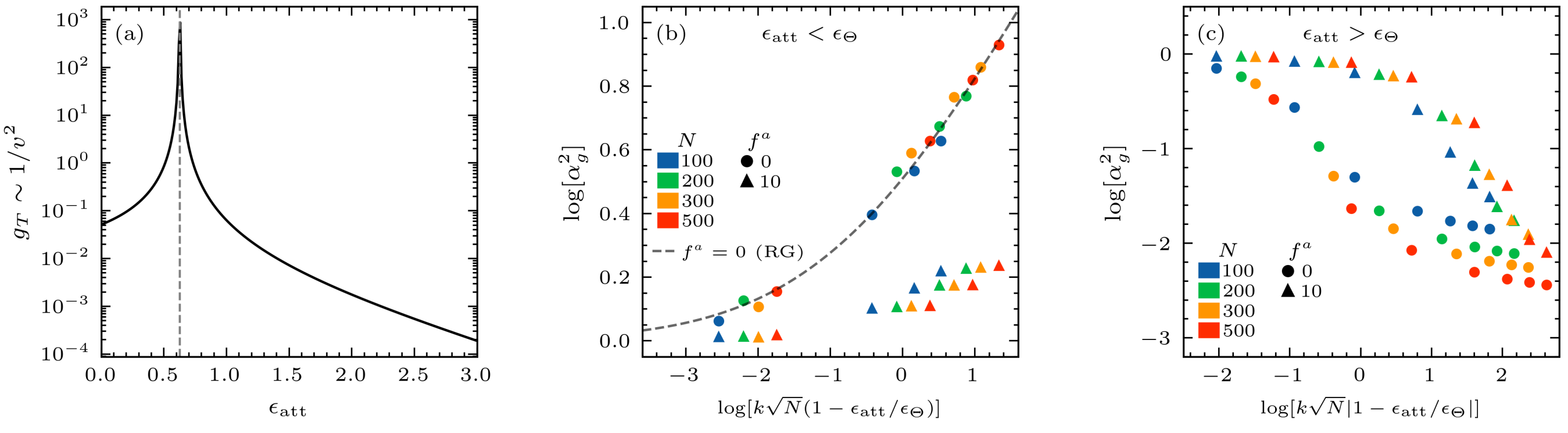


FIG. S1. (a) Thermal blob length scale $g_T$ estimated from the excluded volume parameter $v = \int \mathrm{d}\mathbf{r}\left[1 - e^{-U(\mathbf{r})/k_BT}\right]$, calculated by numerical integration using Eq. S5. (b, c) Squared expansion factor, $\alpha_g^2 \equiv \langle R_g^2\rangle/\langle R_g^2\rangle_\Theta$, as a function of the passive scaling variable $z = k\sqrt{N}(1 - \epsilon_{\rm att}/\epsilon_\Theta)$ for attraction strengths (b) $\epsilon_{\rm att} < \epsilon_\Theta$ and (c) $\epsilon_{\rm att} > \epsilon_\Theta$, at active forces $f^a = 0$ and 10, and various chain lengths $N$. The dashed line represents the theoretical expression for $\alpha_g^2$ for passive polymers obtained from renormalization-group calculations [Eq. S16] [6]

.

## III. PARAMETERS OF THE MODEL

In this work, we vary the solvent quality by changing the depth $\epsilon_{\rm att}$ of an attractive potential [Eq. S4]. Therefore, the $\Theta$ point of the model corresponds to a $\Theta$ attraction strength $\epsilon_\Theta$, and thus

$$z = k\sqrt{N}\left|1 - \frac{\epsilon_{\rm att}}{\epsilon_\Theta}\right|. \tag{S17}$$

We follow the procedure of Ref. [10] to determine the value of $\epsilon_\Theta$ for our passive bead-spring polymer model. According to Eq. S15, a fixed value of the expansion factor $\alpha_g$ should correspond to a fixed value of $z$, and thus to the linear relation $1/\sqrt{N} \sim \epsilon_\Theta - \epsilon_{\rm att}$. Therefore, a plot of $\epsilon_{\rm att}$ as a function of $1/\sqrt{N}$ should yield a straight line whose intercept in the limit $N \to \infty$ gives the $\Theta$ attraction strength $\epsilon_\Theta$. In Fig. S2(a), we show this construction for fixed values of the squared expansion factor $\alpha_g^2$, extracted from our data. Due to finite chain lengths, the different lines intersect before the limit $N \to \infty$ [10]. We read off the intersection point, $\epsilon_\Theta \approx 0.65$, corroborating the value found in Ref. [10] for Monte Carlo simulations.

Following Ref. [8], the chemistry constant $k$ for our model is determined using Eq. S16. Using numerical root-finding, we determine the value of $z$ corresponding to each value of $\alpha_g^2$ obtained from simulations at fixed $N$ and $\epsilon_{\rm att}$. A plot of $z$ as a function of $\sqrt{N}|1 - \epsilon_{\rm att}/\epsilon_\Theta|$ should then yield a straight line with slope equal to the chemistry constant $k$. We find $k \approx 0.17$ [Fig. S2(b)], in agreement with previous works [8].

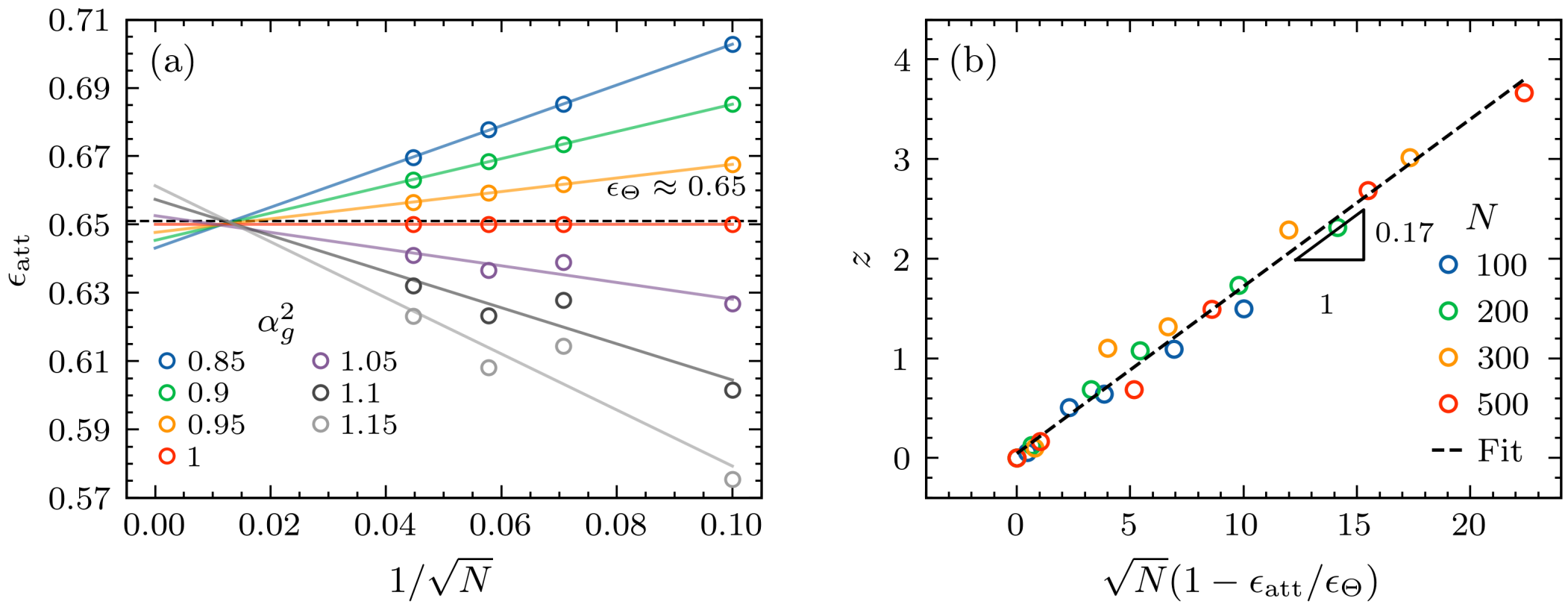


FIG. S2. (a) Attraction strength $\epsilon_{\rm att}$ as a function of $1/\sqrt{N}$ for fixed values of the squared expansion factor, $\alpha_g^2 \equiv \langle R_g^2\rangle/\langle R_g^2\rangle_\Theta$, obtained by linear interpolation between simulation data points. Straight lines correspond to linear fits. The intersection of the curves at $\epsilon_{\rm att} \approx 0.65$ indicates the model's $\Theta$ attraction strength $\epsilon_\Theta \approx 0.65$. (b) Universal scaling variable $z$, obtained from Eq. S16, as a function of $\sqrt{N}(1 - \epsilon_{\rm att}/\epsilon_\Theta)$ for $\epsilon_{\rm att} < \epsilon_\Theta$. The black dashed line corresponds to a linear fit with slope $k \approx 0.17$, corresponding to the model's chemistry constant.

## IV. CONFORMATIONAL PROPERTIES OF ACTIVE POLYMERS

### A. Global conformation

We quantify the typical, global size of polymer chains from their mean-squared radius of gyration

$$\langle R_g^2\rangle = \left\langle \frac{1}{N}\sum_{i=1}^{N}(\mathbf{r}_i - \mathbf{r}_{\rm com})^2 \right\rangle \tag{S18}$$

where $\mathbf{r}_{\rm com} = \frac{1}{N}\sum_{i=1}^{N}\mathbf{r}_i$ is a polymer's center-of-mass, and $\langle\cdot\rangle$ denotes an average of relevant polymer conformations over chains and time. In Fig. S3(a-b), we show the dependence of $\langle R_g^2\rangle$ on the attraction strength $\epsilon_{\rm att}$ for different values of the active force $f^a$ at $N = 100$ and $N = 300$ [for $N = 500$, see main-text Fig. 1(b)]. Below $\epsilon_\Theta$, the effect of increasing activity is similar to that of decreasing the chain length $N$ in the passive limit, $f^a = 0$, as shown in Fig. S3(c).

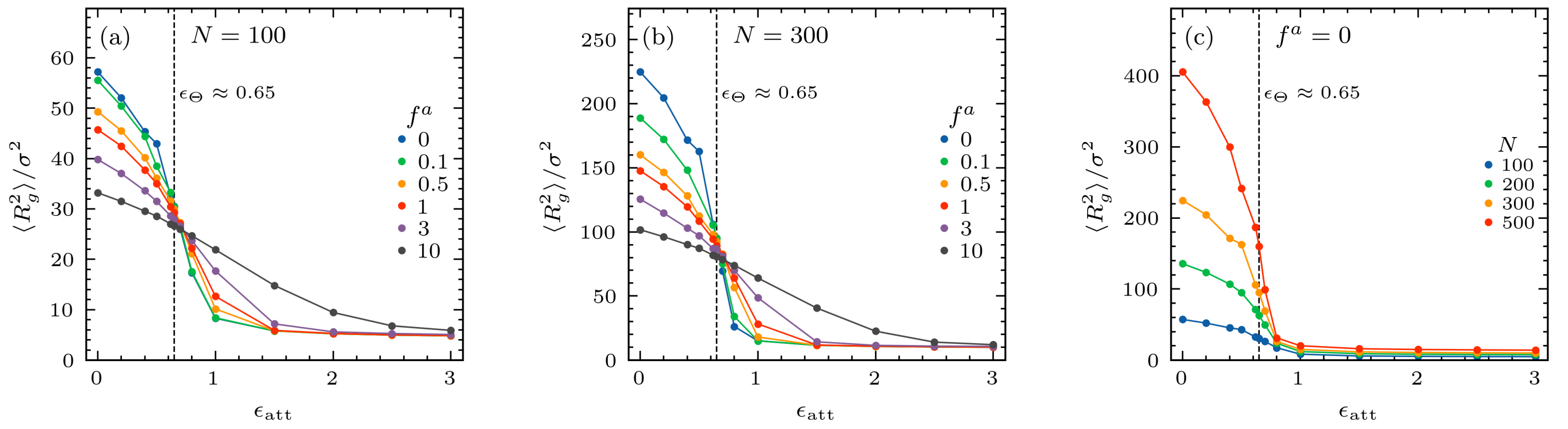


FIG. S3. (a, b) Mean-squared radius of gyration, $\langle R_g^2 \rangle$ [Eq. S18], as a function of the attraction strength $\epsilon_{\mathrm{att}}$ for different active forces $f^a$ at (a) $N = 100$ and (b) $N = 300$. (c) $\langle R_g^2 \rangle$ as a function of $\epsilon_{\mathrm{att}}$ for different chain lengths $N$ at $f^a = 0$.

For all attraction strengths $\epsilon_{\mathrm{att}}$ and active forces $f^a$, the polymer gyration radius follows a power-law dependence on chain length, $\langle R_g^2 \rangle \sim N^{2\nu}$, where $\nu(\epsilon_{\mathrm{att}}, f^a)$ is the generalized global scaling exponent [Fig. S4(a)]. The dependence of $\nu$ on $f^a$ is shown in Fig. 2(a) in the main text. In Fig. S4(b-c), we show $\langle R_g^2 \rangle$ as a function of the active force $f^a$ for various attraction strengths $\epsilon_{\mathrm{att}}$ at chain lengths $N = 100$ and $N = 300$. Remarkably, the effect of activity depends on the position of the attraction strength relative to the $\Theta$ point. For $\epsilon_{\mathrm{att}} < \epsilon_\Theta$, increasing $f^a$ results in chain shrinkage, whereas for $\epsilon_{\mathrm{att}} > \epsilon_\Theta$, activity results in the swelling of compact conformations. At $\epsilon_{\mathrm{att}} = \epsilon_\Theta$, the gyration radius is almost independent of activity, consistent with the discussion in the main text. Also, in the large-activity limit, $\langle R_g^2 \rangle$ converges to the same value regardless of $\epsilon_{\mathrm{att}}$, with this value depending only on the total chain length $N$.

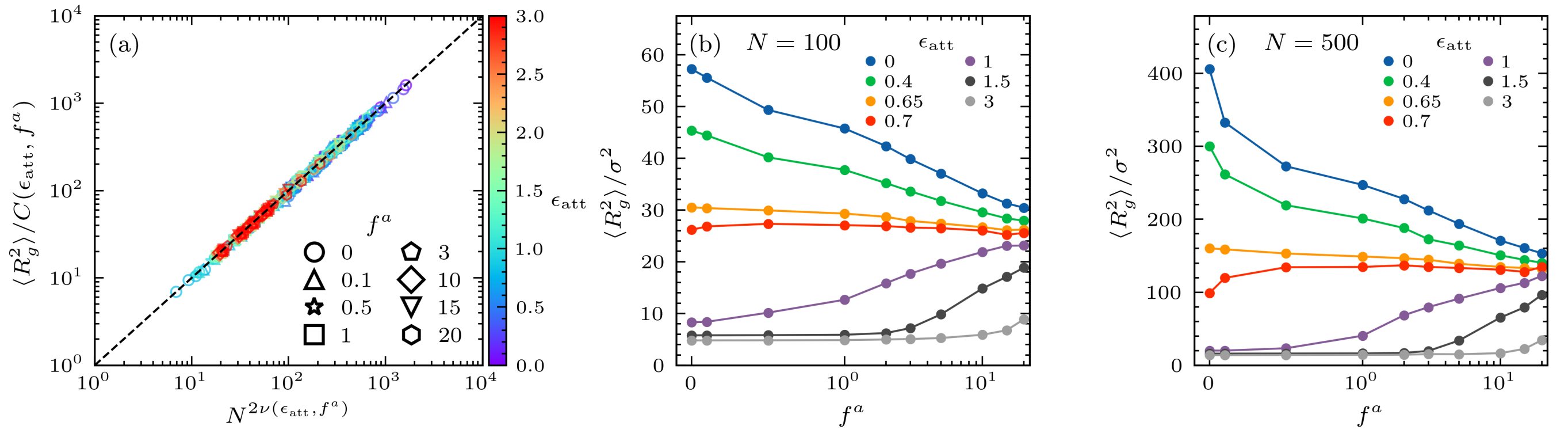


FIG. S4. (a) Normalized mean-squared radius of gyration, $\langle R_g^2 \rangle / C(\epsilon_{\mathrm{att}}, f^a)$ [Eq. S18], as a function of $N^{2\nu(\epsilon_{\mathrm{att}}, f^a)}$. The prefactor $C(\epsilon_{\mathrm{att}}, f^a)$ and scaling exponent $\nu(\epsilon_{\mathrm{att}}, f^a)$ are extracted from power-law fits of $\langle R_g^2 \rangle$ versus $N$. The dashed line corresponds to $y = x$. (b-c) Mean-squared radius of gyration, $\langle R_g^2 \rangle$, as a function of the active force $f^a$ for different attraction strengths $\epsilon_{\mathrm{att}}$ at (b) $N = 100$ and (c) $N = 500$.

## B. Local conformation

To quantify chain conformation at scales smaller than that of the whole polymer, one can compute a local version of Eq. S18 as

$$\langle R_g^2(s) \rangle = \left\langle \frac{1}{N-s} \sum_{i=1}^{N-s} \left[ \frac{1}{s+1} \sum_{j=i}^{i+s} \left( \mathbf{r}_j - \mathbf{r}_{\mathrm{com}}^{(i,s)} \right)^2 \right] \right\rangle \tag{S19}$$

where $1 \le s \le N-1$ is the contour separation along a polymer and $\mathbf{r}_{\mathrm{com}}^{(i,s)} = \frac{1}{s+1} \sum_{j=i}^{i+s} \mathbf{r}_j$. The local scaling exponent can be estimated using central differences as

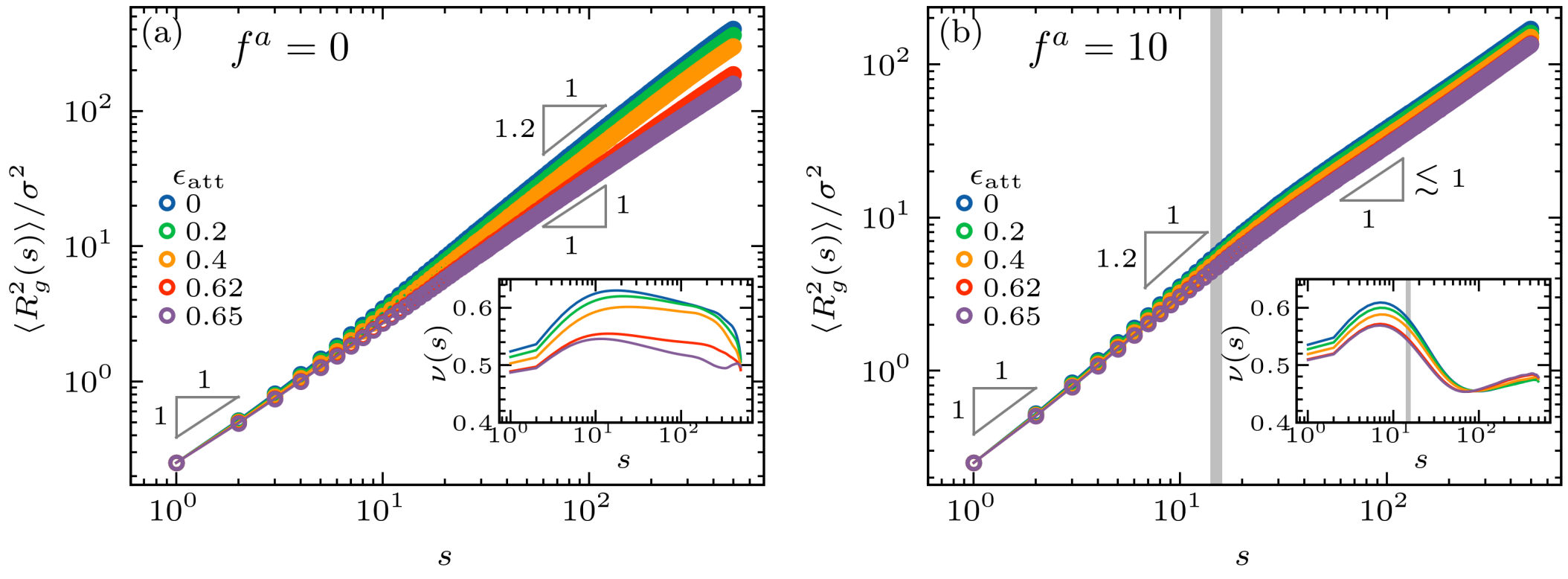


FIG. S5. (a-b) Local radius of gyration, $\langle R_g^2(s)\rangle$ [Eq. S19], as a function of the contour separation $s$ and attraction strength $\epsilon_{\rm att} < \epsilon_\Theta$, at active force $f^a$= (a) 0 and (b) 10, and chain length $N = 500$. Insets: local scaling exponent $\nu(s)$ [Eq. S20]. The shaded areas in (b) indicate the range of positions of the active folding length $s_a = g_a - 1$ [see Sec. V].

$$\nu(s) = \frac{1}{2}\frac{\log\big(\langle R_g^2(s+1)\rangle\big) - \log\big(\langle R_g^2(s-1)\rangle\big)}{\log(s+1) - \log(s-1)}. \tag{S20}$$

In Fig. S5, we plot $\langle R_g^2(s)\rangle$ as a function of the monomer spacing $s$ for attraction strengths $\epsilon_{\rm att} < \epsilon_\Theta$. For passive polymers [panel (a)], the curve at $\epsilon_{\rm att} = 0$ displays a crossover from ideal scaling at small $s$, $\langle R_g^2(s)\rangle \sim s$, to SAW statistics at large $s$, $\langle R_g^2(s)\rangle \sim s^{1.2}$, as expected for self-repelling chains. Upon increasing $\epsilon_{\rm att}$ toward $\epsilon_\Theta$, the large-$s$ behavior gradually evolves toward ideal statistics, reflecting the growth of the thermal blob size $g_T$. This evolution is also visible in the local scaling exponent $\nu(s)$ [inset panel (a)], whose crossover from $\nu_{\rm id}$ at small $s$ to $\nu_{\rm SAW}$ at large $s$ becomes less pronounced as $\epsilon_{\rm att}$ increases. In contrast, $\langle R_g^2(s)\rangle$ for active polymers [panel (b)] exhibits three distinct regimes. From small to intermediate length scales, the usual crossover from ideal to self-avoiding statistics is observed. Similar to the passive case, this crossover is damped as $\epsilon_{\rm att}$ increases [inset of panel (b)], suggesting that thermal effects and the growth of $g_T$ remain relevant in the presence of activity. At contour spacings of the order of the folding length scale $g_a$ [see main-text and Sec. V for definition], the local scaling exponent drops to sub-ideal values, $\nu(s) \lesssim 0.5$, indicating the emergence of large-scale compact structures, in agreement with Ref. [11]. Varying $\epsilon_{\rm att}$ leaves the large-$s$ behavior essentially unchanged, suggesting that the propagation of thermal effects is bounded by $g_a$.

For attraction strengths $\epsilon_{\rm att} > \epsilon_\Theta$ [Fig. S6], the local gyration radius $\langle R_g^2(s)\rangle$ of passive polymers develops a plateau at large $s$, indicative of globular conformations [panel (a)]. We observe that this collapse occurs more gradually and

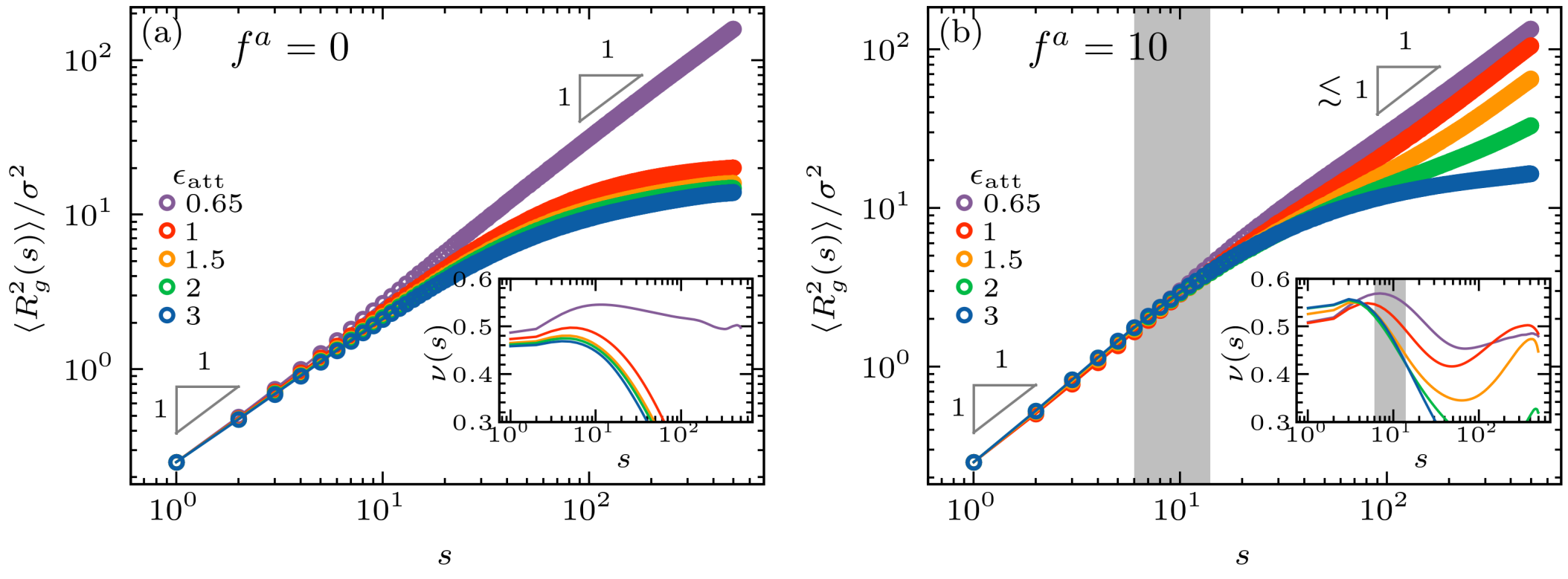


FIG. S6. (a-b) Local radius of gyration, $\langle R_g^2(s)\rangle$ [Eq. S19], as a function of the contour separation $s$ and attraction strength $\epsilon_{\rm att} > \epsilon_\Theta$, at active force $f^a$= (a) 0 and (b) 10, and chain length $N = 500$. Insets: local scaling exponent $\nu(s)$ [Eq. S20]. The shaded areas in (b) indicate the range of positions of the bond-bond correlation function's minimum [Eq. S21], $s_{\rm min} = g_{\rm min} - 1$ [see Sec. V].

at larger attraction strengths for active polymers [panel (b)], consistent with the observations made in the main-text. Interestingly, the position of the minimum of the bond-bond correlation function [Eq. S21] $g_{\mathrm{min}}$, correlates with the crossover between small-$s$ ideal scaling $\nu(s) \approx 1/2$, and the subsequent drop in $\nu(s)$ marking the onset of compact globular structures.

## V. BOND-BOND CORRELATION AND FOLDING LENGTH SCALE

We compute the bond-bond correlation function,

$$C_b(\ell) = \langle \mathbf{b}_i \cdot \mathbf{b}_{i+\ell} \rangle \tag{S21}$$

where $\mathbf{b}_i = \mathbf{r}_{i+1} - \mathbf{r}_i$ and $\ell$ denotes the separation between bonds along the polymer contour. Note that a bond separation $\ell$ corresponds to a contour segment of $g = \ell + 2$ monomers. Following Ref. [11], we define the folding length scale as $\ell_{\mathrm{min}} = \mathrm{argmin}[C_b(\ell)]$, with the corresponding number of monomers given by $g_{\mathrm{min}} = \ell_{\mathrm{min}} + 2$. For $\epsilon_{\mathrm{att}} < \epsilon_\Theta$, folding is purely induced by activity [see Fig. S7 and main-text Fig. 3(a)], and we denote this length scale by $g_a \equiv g_{\mathrm{min}}$. In Fig. S7(b-c), we show $g_a$ as a function $N$ for attraction strength $\epsilon_{\mathrm{att}} = 0$ and $\epsilon_{\mathrm{att}} = 0.65$ ($\Theta$ point). The chain length dependence of $g_a$ is systematically observed as a sublinear power-law scaling, $g_a \sim N^\alpha$, with $\alpha < 1$ slightly dependent on $f^a$ and $\epsilon_{\mathrm{att}}$.

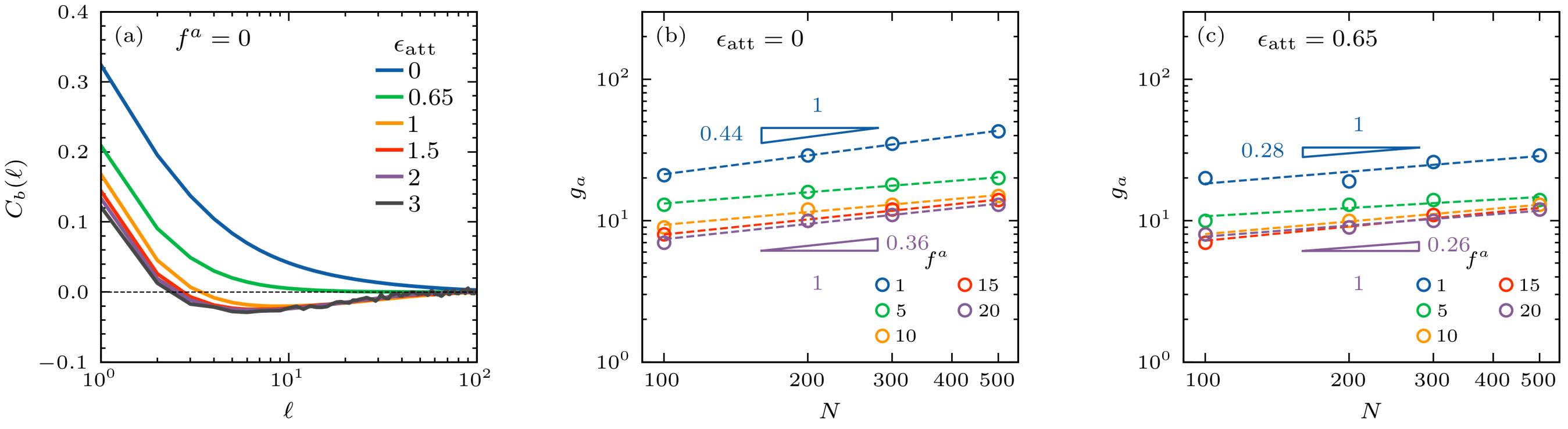


FIG. S7. (a) Bond-bond correlation function $C_b(\ell)$ for attraction strengths $\epsilon_{\mathrm{att}}$ at active force $f^a = 0$ and chain length $N = 500$. (b-c) Active folding length, $g_a$, as a function of chain length $N$ for different active forces $f^a$ at attraction strengths $\epsilon_{\mathrm{att}} =$ (b) 0 and (c) 0.65 ($\Theta$ point). Dashed lines show power-law fits, $g_a \sim N^\alpha$.

## VI. ACTIVE BLOB PICTURE

The number of thermal blobs within a polymer, $N/g_T$ [Eq. S13], fails to collapse the expansion factor of active polymers, as shown in Fig. S1(b-c). In this section, we derive a new scaling variable for the collapse of active polymers for attraction strengths $\epsilon_{\mathrm{att}} < \epsilon_\Theta$, following the discussion in the main-text. In Fig. S8, we plot $\langle R_g^2 \rangle / \xi_a^2$, where $\xi_a^2 = \langle R_g^2(g_a) \rangle$, as a function of $N/g_a$ below $\epsilon_\Theta$, and $\langle R_g^2 \rangle / \xi_{\mathrm{min}}^2$ as a function of $N/g_{\mathrm{min}}$ above $\epsilon_\Theta$. For $\epsilon_{\mathrm{att}} \leq \epsilon_\Theta$ [panels (a) and (b)], power-law scaling is observed and curves for distinct active forces collapse well onto each other, up to minor prefactor and exponent variations. The ansatz

$$\langle R_g^2 \rangle \sim \xi_a^2 \left( \frac{N}{g_a} \right)^{2\nu_a} \tag{S22}$$

is verified, with $\nu_a \lesssim 1/2$ nearly independent of attraction strength and activity [see Fig. S8(a-b)]. For attraction strengths $\epsilon_{\mathrm{att}} > \epsilon_\Theta$, curves for distinct activities split apart, with those at low $f^a$ shifting downward while those at large $f^a$ remain close. This indicates that strong attractive interactions impact the conformation of active polymers at low activity, whereas polymers at high activity remain largely independent of solvent quality.

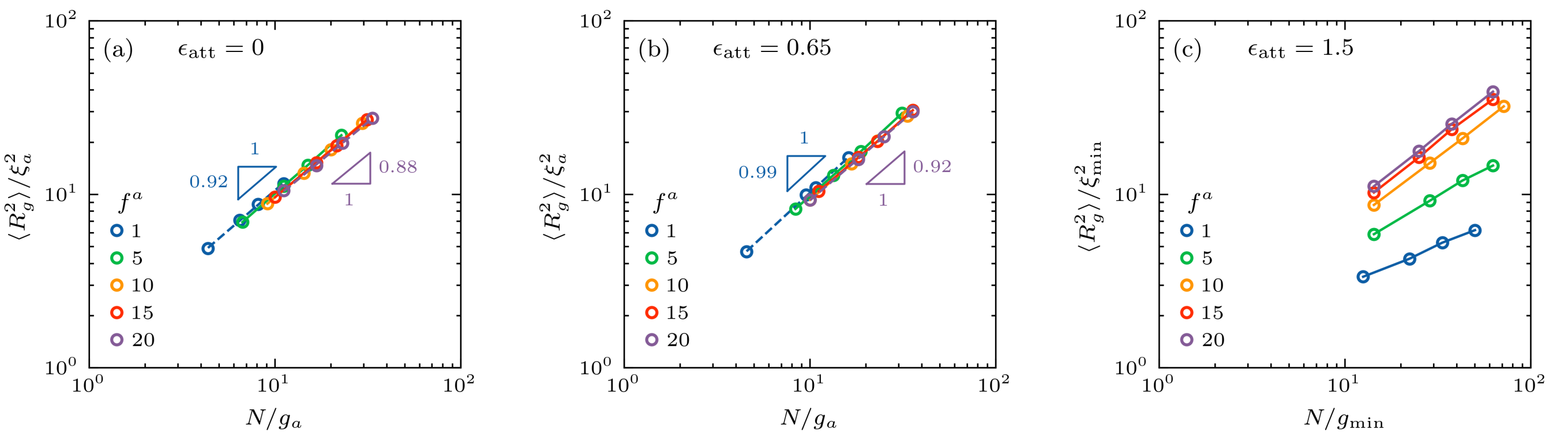


FIG. S8. Normalized mean-squared radius of gyration $\langle R_g^2 \rangle / \xi_a^2$ [Eq. S18], with $\xi_a^2 = \langle R_g^2(g_a) \rangle$ [Eq. S19], as a function of $N/g_a$ for attraction strengths $\epsilon_{\rm att} =$ (a) 0 and (b) 0.65 ($\Theta$ condition), at different active forces $f^a$. The triangles indicate the slopes on a log-log scale, corresponding to the exponent $\nu_a$ [Eq. S22]. (c) Same for $\epsilon_{\rm att} = 1.5$, but using $\xi_{\rm min}^2 = \langle R_g^2(g_{\rm min}) \rangle$ and $g_{\rm min}$, since the folding length is no longer purely activity-induced above $\epsilon_\Theta \approx 0.65$.

In the main text, we established that for attraction strengths $\epsilon_{\rm att} \leq \epsilon_\Theta$, the thermal blob ansatz

$$\xi_a^2 \sim \xi_T^2 \left( \frac{g_a}{g_T} \right)^{2\gamma} \tag{S23}$$

holds true [see main-text Fig. 3(c)]. Combining Eq. S22 and Eq. S23, we have

$$\langle R_g^2 \rangle \sim \xi_T^2 \left( \frac{g_a}{g_T} \right)^{2\gamma} \left( \frac{N}{g_a} \right)^{2\nu_a} \sim g_T^{1-2\gamma} g_a^{2[\gamma - \nu_a]} N^{2\nu_a}. \tag{S24}$$

It follows that the expansion factor of tangentially-driven active polymers should writes

$$\alpha_g^2(\epsilon_{\rm att}, N, f^a) \equiv \frac{\langle R_g^2 \rangle}{\langle R_g^2 \rangle_\Theta} \sim \frac{g_T^{1-2\gamma} g_a(\epsilon_{\rm att}, N, f^a)^{2[\gamma - \nu_a]}}{g_a(\epsilon_\Theta, N, f^a)^{1-2\nu_a}} \tag{S25}$$

$$\sim \left( \frac{g_a(\epsilon_{\rm att}, N, f^a)}{g_a(\epsilon_\Theta, N, f^a)} \right)^{1-2\nu_a} \left( \frac{g_a(\epsilon_{\rm att}, N, f^a)}{g_T} \right)^{2\gamma - 1}. \tag{S26}$$

where the first term is approximately unity, as $g_a(\epsilon_{\rm att}) \approx g_a(\epsilon_\Theta)$ and $1 - 2\nu_a \approx 0$ [Fig. S8]. From Eq. S26, we read-off the active scaling variable $g_a/g_T$, such that generally on expects

$$\alpha_g \sim \left( \frac{g_a}{g_T} \right)^{\gamma - 1/2}, \quad \text{or} \quad \alpha_g \sim z_a^{2\gamma - 1}, \tag{S27}$$

where $z_a = k\sqrt{g_a}(1 - \epsilon_{\rm att}/\epsilon_\Theta)$, which appears as the active counterpart of Eq. S14. Therefore, the expansion factor of a tangentially driven active polymer is determined by the number of thermal blobs, containing $g_T$ monomers, within an active blob containing $g_a$ monomers. It is noteworthy that an activity-dependent prefactor should be expected in Eq. S27, as not only $g_a$ but also the magnitude of the bond-bond correlation, $|C_b(\ell_a)|$, should enter into the picture. In the limit of finite-chain size, we expect this scaling to generalize to

$$\alpha_g \sim F_a\left(z_a\right) \tag{S28}$$

where $F_a$ is a crossover scaling-function that, *a priori*, depends on the active force $f^a$. In the main text, we show that this prediction is verified and that the expansion factors collapse onto the passive polymer master curve, Eq. S16, upon introducing the activity-dependent metric factor $\mathcal{A}(f^a) = k^2(f^a)^{-0.72}$ [see main-text Fig. 4(a-b)]. We rationalize this prefactor by showing that $\mathcal{A}(f^a)$ scales as $1/|C_b(\ell_{\rm min})|$ [see main-text Fig. 4(c)]. In Fig. S9, we show $\alpha_g^2$ as a function of $0.07k\sqrt{g_a/|C_b(\ell_{\rm min})|}(1 - \epsilon_{\rm att}/\epsilon_\Theta)$, which explicitly accounts for bond-bond correlations in the scaling variable. We observe that this scaling variable also yields a satisfactory collapse of $\alpha_g^2$ for different active forces $f^a$ onto the passive master curve.

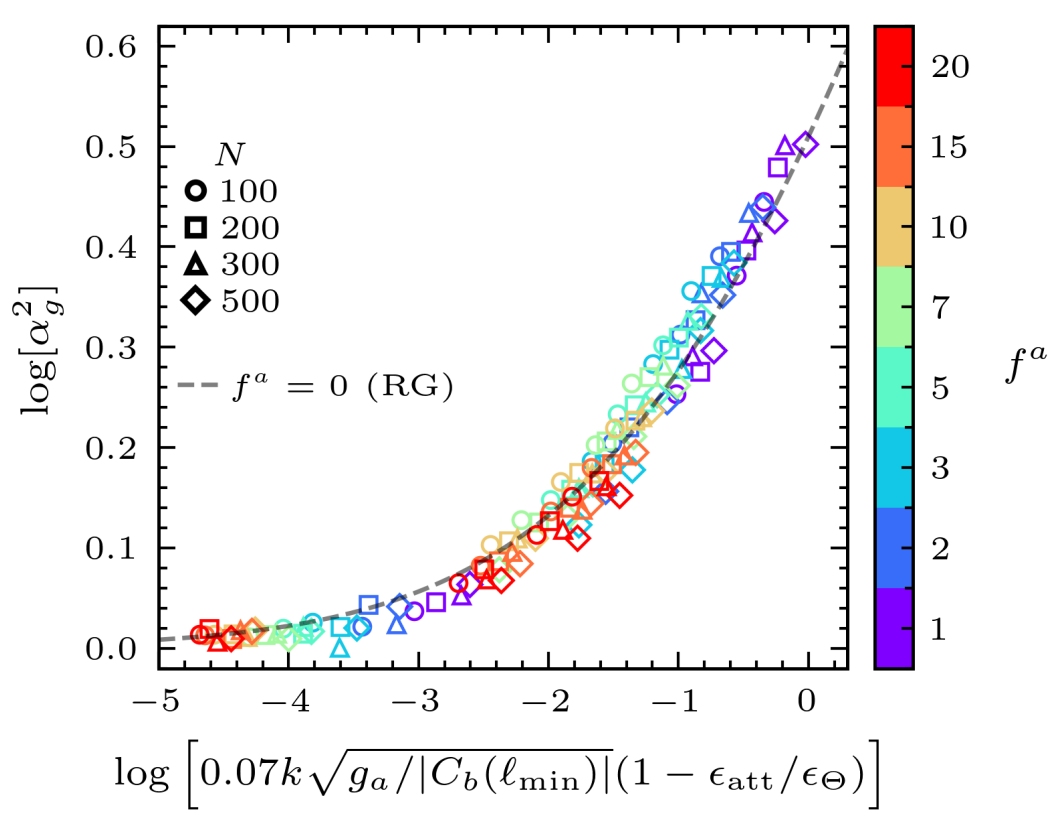


FIG. S9. Squared expansion factor, $\alpha_g^2 \equiv \langle R_g^2\rangle/\langle R_g^2\rangle_\Theta$, as a function of $0.07k\sqrt{g_a/|C_b(\ell_{\min})|}(1-\epsilon_{\rm att}/\epsilon_\Theta)$, where $k \approx 0.17$ is the chemistry constant of the model [see Sec. III]. The dashed line represents the theoretical expression for $\alpha_g^2$ for passive polymers obtained from renormalization-group calculations [Eq. S16] [6].

## VII. CONTACT PROPERTIES

For attraction strengths $\epsilon_{\rm att} > \epsilon_\Theta$, active polymers collapse into compact, globular structures. To quantify the effect of activity on the internal structure of the globule, we compute the contact probability between two monomers,

$$p_c(s) = \langle \mathrm{H}(r_{\rm cut} - r_{i,i+s})\rangle, \tag{S29}$$

where H is the Heaviside step function and $r_{\rm cut}$ is set to $1.5\sigma$. In Fig. S10(a-b), we plot $p_c(s)$ as a function of $s$ for polymers of length $N = 500$. In the fully repulsive limit [panel (a)], the contact probability is independent of the active force $f^a$ at small $s$, consistent with the fact that activity does not affect the short length scales. At large contour separations, $p_c(s)$ shifts upward with increasing $f^a$, indicating enhanced monomer collisions, in agreement with the onset of compact folded structures. Conversely, at $\epsilon_{\rm att} = 3$ [panel (b)], the contact probability remains independent of $f^a$, indicating that activity does not alter the underlying globular structure.

To investigate the dynamics of active polymer globules, we define a contact time autocorrelation function (TACF) by assigning a contact variable $n_{ij}(t)$ for every pair of monomer $i$ and $j$ inside a chain. This variable is set to 1 if particles $i$ and $j$ are withtin a cut-off distance, chosen at $1.5\sigma$, and 0 otherwise. The contact TACF is then calculated

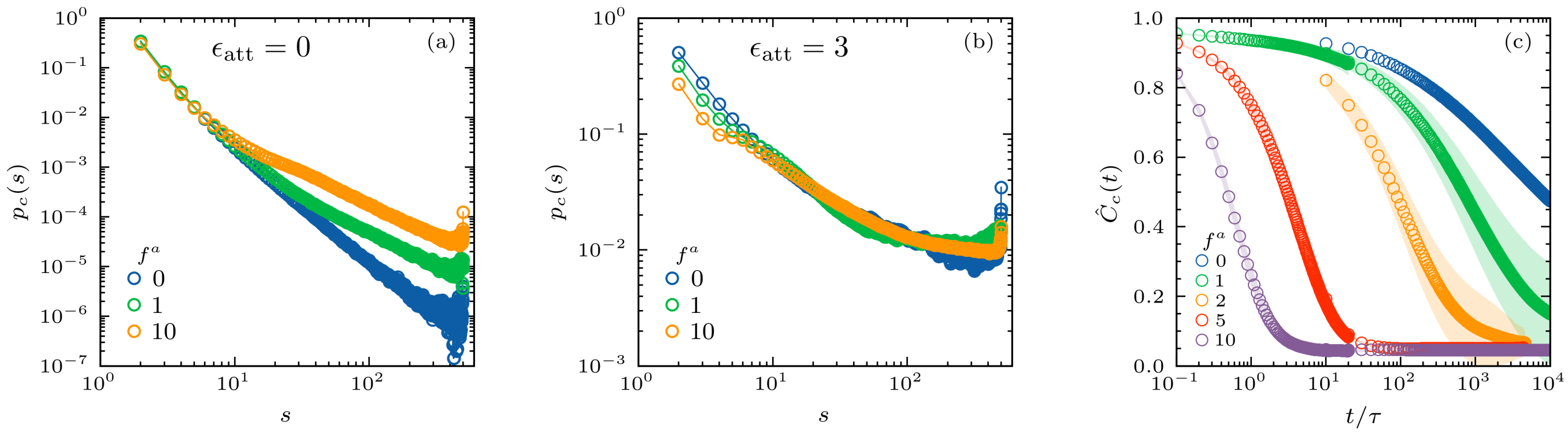


FIG. S10. Contact probability $p_c(s)$ [Eq. S29] as a function of the monomer spacing $s$ at attraction strengths $\epsilon_{\rm att} =$ (a) 0 and (b) 3 for different active forces $f^a$ and chain length $N = 500$. (c) Contact-time autocorrelation function $\hat{C}_c(t)$ [Eq. S30] at attraction strength $\epsilon_{\rm att} = 3$ for different active forces $f^a$ and chain length $N = 500$. Shaded areas indicate the variance across polymer chains.

as

$$\hat{C}_c(t) = \frac{\langle \sum_{i\neq j} n_{ij}(t')n_{ij}(t'+t)\rangle}{\langle \sum_{i\neq j} n_{ij}(t')^2\rangle}. \tag{S30}$$

The contact TACF is shown as a function of time in Fig. S10(c) for polymers of length $N = 500$. Interestingly, despite their compact nature, active globules are highly dynamic, with $\hat{C}_c(t)$ exhibiting a faster decay as $f^a$ increases, indicating the emergence of reconfigurable globular structures.

## VIII. CLUSTERING ANALYSIS

The structural changes occurring during the collapse of passive and active polymers are analyzed by monitoring the number of clusters along the chains in the steady state. Following Paul *et al.* [12], we compute the number of neighboring monomers $n_i$ around monomer $i$ within a cut-off distance $r_{\rm cut}$,

$$n_i(t) = \sum_{j\neq i} \mathrm{H}(r_{\rm cut} - r_{ij}) \tag{S31}$$

where H is the Heaviside step function. A monomer is considered *clustered* if it has a sufficiently high local neighbor density $\rho_{\rm min}$, *i.e.*, if $n_i(t) > n_{\rm min}$ within a spherical volume of radius $r_{\rm cut}$. Finally, we obtain the number of clusters $n_c(t)$ along a chain by merging each monomer's neighbor list using a union-find approach [13]. The parameters were chosen to be consistent with the visual inspection of the collapse of passive chains ($f^a = 0$). The local threshold density was set to $\rho_{\rm min} \approx 0.3$ ($\rho_{\rm coil} < \rho_{\rm min} < \rho_{\rm glob}$), with $n_{\rm min} = 20$ and $r_{\rm cut} = 2.5\sigma$, such that $\langle n_c\rangle = 0$ at $\epsilon_{\rm att} = 0$ and $\langle n_c\rangle = 1$ at $\epsilon_{\rm att} = 3$.

Structural states are identified from the mean number of clusters $\langle n_c\rangle$: coil for $\langle n_c\rangle < 1-\epsilon$, necklace for $\langle n_c\rangle > 1+\epsilon$, and globule for $|\langle n_c\rangle - 1| \leq \epsilon$. We set $\epsilon = 0.05$ to match visual inspection. See main-text Fig. 6.

---



\end{document}